\documentclass[aps, prd,superscriptaddress,preprintnumbers, onecolumn,floatfix,nofootinbib]{revtex4-1}
\usepackage{amsmath,amssymb,graphicx,tabularx,url,multirow,hyperref}

\newenvironment{itemize*}
  {\begin{itemize}
    \setlength{\itemsep}{0pt}
    \setlength{\parskip}{0pt}}
  {\end{itemize}}

\newenvironment{enumerate*}
  {\begin{enumerate}
    \setlength{\itemsep}{0pt}
    \setlength{\parskip}{0pt}}
  {\end{enumerate}}

\newenvironment{description*}
  {\begin{description}
    \setlength{\itemsep}{0pt}
    \setlength{\parskip}{0pt}}
  {\end{description}}

\def\gsim{\lower0.5ex\hbox{$\:\buildrel >\over\sim\:$}}
\def\lsim{\lower0.5ex\hbox{$\:\buildrel <\over\sim\:$}}

\newcommand{\be}{\begin{equation}}
\newcommand{\ee}{\end{equation}}
\newcommand{\bea}{\begin{eqnarray}}
\newcommand{\eea}{\end{eqnarray}}

\newcommand{\nbox}{{\,\lower0.9pt\vbox{\hrule \hbox{\vrule height 0.2 cm
\hskip 0.2 cm \vrule height 0.2 cm}\hrule}\,}}

\def\met {{\not\!\! E_T}}
\def\zp {{Z^\prime}}

\begin{document}

\preprint{CETUP2015-016}

\title{Searches for Dark Matter at the LHC: A Multivariate Analysis in the Mono-$Z$ Channel}

\author{Alexandre Alves}
\affiliation{Departamento de Ci\^encias Exatas e da Terra,
Universidade Federal de S\~ao Paulo, 09972-270, Diadema-SP, Brasil}

\author{Kuver Sinha}
\affiliation{Department of Physics, Syracuse University, Syracuse, NY 13244, USA}


\date{\today}

\begin{abstract}
We study dark matter (DM) production in the mono-Z channel at the 13 TeV LHC both in an effective field theory framework as well as in simplified models with vector mediators, using a multivariate analysis. For DM-quark effective operators with scalar, vector, and tensor couplings and DM mass of 100 GeV, the 5$\sigma$ reach in the DM interaction scale $\Lambda$ is around 2, 1, and 3 TeV, respectively, for 3 ab$^{-1}$ and assuming a 5\% systematic uncertainty on the total background normalization. For simplified models with leptophobic vector mediators, the 5$\sigma$ reach for the mass of the mediator is 1.7 TeV also assuming a 5\% systematics and 3 ab$^{-1}$ of integrated luminosity. The reach for the dark matter interaction scale obtained with the multivariate analysis using a likelihood function discriminant is at least twice as high as that obtained from a simple cut and count analysis, once systematics on the background normalization larger than a few percent are taken into account. Moreover, the reach is much more stable against degradation due these systematic uncertainties. 
\end{abstract}

\pacs{}
\maketitle

\section{Introduction}

The particle nature of dark matter (DM) is one of the most important unsolved questions in physics, and the focus of immense experimental and theoretical endeavors. One of the most promising arenas for such endeavors is the study of DM in colliders.

Approaches to study the dark matter parameter space include effective operators \cite{Goodman:2010ku} and simplified models \cite{Abdallah:2014hon}, with a mono-$X$ observational signature, where $X$ may denote monojet \cite{Beltran:2010ww}, mono-photon \cite{Gershtein:2008bf}, mono-$Z$ \cite{Petriello:2008pu, Carpenter:2012rg}, mono-$W$ \cite{Bai:2012xg}, mono-Higgs, mono-$b$ \cite{Lin:2013sca}, or mono-top \cite{Lin:2013sca}. In studies with effective operators, limits on the cut-off scale $\Lambda$ of the low energy DM effective field theory have been reported.

The purpose of the current paper is to explore the reach at 13 TeV for the effective interaction scale between DM and quarks as well as electroweak bosons. Moreover, we also study simplified models with DM-quark interactions mediated by leptophobic vector mediators. We explore the mass reach for such mediators. We choose the mono-$Z$ channel for our study, since, compared to monojet, it is expected that background uncertainties should scale more favorably for the mono-$Z$ signature.  In mono-$Z$ studies, a pair of charged leptons that reconstruct to a $Z$ boson recoil against the missing transverse momentum carried by dark matter. The dominant background in this case is constituted by SM processes $ZZ \rightarrow \ell \ell \nu \nu$. Using ATLAS bounds on $ZZ \rightarrow \ell \ell \nu \nu$ \cite{Aad:2012awa}, constraints were put on the dark matter effective scale $\Lambda$ in  \cite{Carpenter:2012rg}. 

We will be especially careful to incorporate the effects of systematic uncertainties, which are expected to become increasingly important in the current run of the LHC. Our strategy is to obtain a set of kinematic variables that can discriminate between signal and background, and construct a multivariate likelihood function on which cuts can be imposed. Broadly, we find that the reach for the DM interaction scale $\Lambda$ obtained with the multivariate analysis using a likelihood function is at least twice as high as that obtained from a simple cut and count analysis, when systematic uncertainties greater than a few percent are taken into account. Moreover, the reach is much more stable against degradation due to systematics in the total background normalization. 

The rest of the paper is structured as follows. In Section \ref{DMEFTandZprime}, we first discuss both DM effective operators and simplified models with vector mediators. In Section \ref{variables}, we discuss the kinematic variables used in the construction of the multivariate discriminant. In Section \ref{systuncert}, we discuss systematic uncertainties. In Section \ref{MVA}, we discuss the construction of the multivariate (MVA) discriminant. In Section \ref{EFTresults}, we present the main results of our multivariate study for the DM effective operators, while in Section \ref{ZPresults}, we present the results for the $\zp$ model. We end with our conclusions.


\section{Background: Effective Operators and Vector Mediators} \label{DMEFTandZprime}

In this Section, we describe the theoretical background of our study. We first discuss DM effective operators, and then turn to simplified models with vector mediators.

\subsection{Effective Operators}

Effective field theories for dark matter interacting primarily with SM quarks have been considered in Refs.~\cite{Beltran:2008xg,Shepherd:2009sa,Cao:2009uw,Beltran:2010ww,Goodman:2010yf,Bai:2010hh,Goodman:2010ku,Rajaraman:2011wf,Fox:2011pm,Cheung:2012gi}.  

We consider the following interactions
\bea
 &  \sum_q & \left\{
\frac{1}{\Lambda_{\rm \tilde{D}1}^2} \bar{q} q~ \bar{\chi} \chi
+ \frac{1}{\Lambda_{\rm D8}^2}  \bar{q} \gamma^\mu \gamma_5 q~
\bar{\chi} \gamma_\mu \gamma_5 \chi \right.
\nonumber \\ & & \left.
+ \frac{1}{\Lambda_{\rm D5}^2}  \bar{q} \gamma^\mu q~ \bar{\chi} \gamma_\mu \chi
+ \frac{1}{\Lambda_{\rm D9}^2}  \bar{q} \sigma^{\mu \nu} q~
\bar{\chi} \sigma_{\mu \nu} \chi
\right\}
\label{eq:EFTq}
\eea
In the above, $\chi$ denotes a Dirac fermion that we assume to be the DM particle. The SM quarks are denoted by $q$. $\tilde{\hbox{D}}$1, D5, D8, and D9 denote effective DM-$q$ operators with scalar, vector, axial-vector, and tensor couplings. To simplify the analysis, we will assume that only one type of operator is dominant at one time, with the others decoupled. For the scalar operator $\tilde{\hbox{D}}$1 we assume an underlying model with a scalar mediator $S$ with democratic couplings to all quark flavors. A Higgs portal model, on the other hand, leads to quark mass suppressed couplings which cannot be probed by the LHC.

We also consider two operators considered in \cite{Carpenter:2012rg}
\be
L_{5} \, \equiv \, \frac{1}{\Lambda^3_5} \overline{\chi} \chi (D_\mu H)^{\dagger} D^{\mu} H
\ee
which arises at dimension 5, and
\be
L_{7} \, \equiv \, \frac{1}{\Lambda^3_7} \overline{\chi} \chi \sum_i k_1 F^{\mu \nu}_i F^i_{\mu \nu} \,\,,
\ee
which arises at dimension 7. Here, $F_i$ with $i = 1,2,3$ are the field strengths for the SM gauge groups.


In \cite{Carpenter:2012rg}, the CLs method was used to put limits on the cross section for new physics (and hence the scales $\Lambda$ for the different operators) based on existing ATLAS results. An upper limit on the number of events coming from new physics was obtained from the expected number of background events with uncertainties. The $90\%$ CL exclusion bounds on the scales $\Lambda$ were given with 7 TeV data, and it was found that for some types of operators, scales of the order of 100 GeV - 1 TeV could be probed.

\subsection{Vector Mediators}

We now turn to simplified models with a leptophobic vector mediator. The Lagrangian is
\begin{eqnarray}\label{lgrg1}
{\cal L} &=& -\frac{g}{2c_W} \left[\sum_{i} g_f\bar{q_i}\gamma^{\mu}(1 - \gamma_{5}) q_i Z'_{\mu}
+ g_\chi \bar{\chi} \gamma^{\mu} (1 -\gamma_{5}) \chi \right]  Z'_{\mu}\ ,
\label{eq1}
\end{eqnarray}
 where the $q_i$ are SM quarks, $g_f$ and $g_\chi$ parametrize the coupling between quarks and the vector mediator $Z^{\prime}$, and between the DM candidate $\chi$ and the $\zp$, respectively, in terms of the SM coupling $\frac{g}{2c_W}$, where $g$ is the $SU(2)_L$ coupling and $c_W$ is the cosine of the Weinberg angle. In our simulations we choose $g_f=g_\chi=1$ and $m_\chi = 100$ GeV as a benchmark point. 

 For the benchmark point, the branching ratios of a 500 GeV $\zp$ to dark matter, jets and top quark pairs are 5.2\%, 80.2\% and 14.7\%, respectively, and remain unchanged for larger masses. The ratio of the total width to $m_{\zp}$ is less than 0.1 for masses up to 4 TeV.
%
%

Collider studies of the $\zp$ model have been performed by many authors in recent years, starting from the work of \cite{Bai:2010hh}. Bounds from direct and indirect dark matter detection and from colliders for a leptophobic $\zp$ portal model were derived in~\cite{Alves:2013tqa}. Overall, for couplings of SM size and a 100 GeV DM, strong bounds from LUX and dijet searches at the LHC 8TeV exclude a $\zp$ portal up to $m_{\zp} \, \sim \, 2.8$ TeV. However, these bounds drop to $m_{\zp} \, \sim \, 2$ TeV for a 50\% suppressed couplings to quarks or heavier DM. On the other hand, the branching ratio of the $\zp$ to DM is mildly dependent on $m_\chi$ until it gets very close to the decay threshold, but increases very fast as the couplings between $\zp$ and quarks get smaller. As a result, for a 500 GeV dark matter and 50\% suppressed couplings to quarks, an 1 TeV $\zp$, for example, is not excluded by any experiment and is able to reproduce the observed DM relic density~\cite{Alves:2013tqa}.

\section{Kinematic Variables for Multivariate Analysis} \label{variables}

In this Section, we describe the kinematic variables that will be used in our multivariate analysis. In the construction of a discriminant with a higher performance level than simple event counting, at least in certain regions of phase space, the information from the shape of the kinematic distributions is crucial. Combining this information in a likelihood function is one of the most simple, robust and straightforward means to build a multivariate discriminant between signal and background events. 

In order to discriminate a dark matter signal at the LHC, we study the process 
\be
pp \, \rightarrow \, Z (\rightarrow \ell^+ \ell^-) \chi \overline{\chi} \,\,, 
\ee
where the DM $\chi$ interacts with quarks via operators $\tilde{\hbox{D}}$1, D5, D8, and D9 and directly with gauge bosons via operators $L_5$ and $L_7$. We also considered a process where dark matter is produced through the production and decay of a new leptophobic massive gauge boson $\zp$ 
\be
pp \, \rightarrow \, Z (\rightarrow \ell^+ \ell^-) Z^\prime (\rightarrow \chi \overline{\chi}) \,\,, 
\ee

The DM effective operators and the $\zp$ simplified model were implemented in \texttt{FeynRules}~\cite{Alloul:2013bka}. Signal and background samples for $\ell^+\ell^-+\met$ up to one extra jet are generated with \texttt{Madgraph}~\cite{Alwall:2011uj} followed by the parton showering and hadronization with \texttt{Pythia}~\cite{Sjostrand:2006za} and the detector simulation using \texttt{Delphes}~\cite{deFavereau:2013fsa}. 
The hard (\texttt{MadGraph}) and soft jet (\texttt{Pythia}) regimes are matched according to the MLM~\cite{Mangano:2006rw} prescription at the matching scale $Q_{cut}=20$ GeV for the EFT operators and 40 GeV for the heavy $\zp$ bosons\footnote{Although we do not use it in the curent work, we note that NLO cross sections for the signal have been computed at the parton level and implemented in MCFM by \cite{Fox:2012ru}, and in \texttt{POWHEG} by \cite{Haisch:2013ata}.}. In Table~(\ref{tab:xsec}) we display the production cross section for each operator for various dark matter masses keeping $\Lambda_i = 1$~TeV for 13 TeV LHC and for the $\zp$ production and decay to dark matter in association to a SM $Z$ boson keeping $g_f=g_\chi=1$ and $m_\chi=100$ GeV fixed. 

\begin{table}
\caption{Production cross sections (in fb) for pair production of DM in
  association with a $Z$ boson and one extra QCD jet at LHC 13 TeV:
  $pp\rightarrow Z\chi\bar{\chi} + j$. In the left part of the table we display the rates for the EFT operators considered in this work for $\Lambda_i = 1$~TeV as a function of the DM mass. In the right part, the rates for the leptophobic $\zp$ model are given as a function of the $\zp$ mass for $g_f=g_\chi=1$ and $m_\chi=100$ GeV.}
\label{tab:xsec}
\begin{tabular}{lrrrrcr|cr}
\hline
$m_{\chi}$ (GeV) &  ~~~~$\tilde{\hbox{D}}$1 & ~~~D5 & ~~~D8 & ~~~D9 & ~~~Dim-5($\times 10^{-4}$) & ~~~Dim-7 & ~~~$M_\zp$ (GeV) & ~~~$\zp$ \\
\hline
  $\le 10$ &	275 & 47 & 47 & 1219 & $2.5$ & 43 & 500 & 144 \\
~~~100   &	257 & 46 & 43 & 1204 & $1.2$  & 39 & 1000 & 21.1 \\
~~~200   &	218 & 42 & 34 & 1058 & $0.59$  & 31 & 2000 & 1.50 \\
~~~400   &	142 & 32 & 20 &  782 & $0.18$  & 19 & 3000 & 0.18 \\
~~~1000  &       26 &  8 &  3 &  205 & $0.08$ &  3 & 4000 & 0.03 \\
\hline
\end{tabular}
\end{table}

The $ZZ \rightarrow \ell \ell \chi \overline{\chi}$ events are characterized by large $\met$ and the presence of two high $p_T$ isolated electrons or muons. The following preselection cuts were imposed, inspired by the ATLAS study of $ZZ \rightarrow \ell^+ \ell^- + \met$ \cite{Aad:2012awa}:
\begin{itemize}
\item two same-flavor opposite-sign electrons or muons, each with $p_{\rm
    T}^{\ell} > 20$ GeV in $|\eta^{\ell}|<2.5$ and dilepton invariant mass close to the $Z$ boson mass:
  $m_{\ell\ell} \in [76, 106]$ GeV; $\Delta R_{\ell\ell} > 0.3$
\item  veto jets with $p^j_{\rm T} >$ 25 GeV and
  $|\eta^j|<$4.5.
\end{itemize}
We also investigated an event selection without imposing the jet veto but keeping the $p_T$, $\eta$, and $m_{\ell\ell}$ requirements.

At the 13 TeV LHC, the dominant SM backgrounds after the $Z$ window selection consists of $ZZ + j$ and $t\overline{t}$ events, with a sub-dominant contribution from $WW, Wt, WZ, \tau^+\tau^-$. The $Wt$ and $\tau^+\tau^-$ backgrounds are negligible after we demand a MET cut:
\begin{itemize}
\item $\met > 100,150,\cdots , 500$ GeV
\end{itemize}

We chose the following kinematic variables to better classify the signal and background events both in the EFT as well as the simplified $\zp$ model frameworks:
\begin{itemize}

\item Missing energy $\met$ is expected to be the kinematic variable that offers one of the best avenues of discerning signal and background. However, this variable is used as a cut variable in order to reduce the number of background events but not to build the likelihood functions. At the LHC, the transverse missing energy serves as a trigger to new physics and a cut on $\met$ is always necessary. In any case, $\met$ is correlated to other variables which we actually use to construct the multivariate discriminant of our signal models and the backgrounds.


\item The product $\met \, \times \, \cos\left(\Delta\phi(\vec{E}_T^{miss},\vec{p}_T^Z)\right)$, where $\Delta \phi$ is the angle between the two dimensional vector $\vec{E}_T^{miss}$ and the transverse momentum $\vec{p}_T^Z$ of the $Z$ boson candidate. This serves as a measure of axial-$\met$, which is defined as the projection of $\vec{E}_T^{miss}$ along the direction opposite to the $Z$ candidate~\cite{Aad:2012awa}. This variable has a great potential to discern between the various DM operators as suggested in~\cite{Carpenter:2012rg}.

\item The fractional $p_T$ difference $|\met - p_T^Z|/p_T^Z$~\cite{Aad:2012awa}.

\item The azimuthal separation of the two leptons $\Delta\phi(\ell^+,\ell^-)$.

\item $\alpha_T \, = \, E_T(\ell_2)/M_T$, where $E_{T2}$ is the transverse energy of the softest lepton of the $\ell^+\ell^-$ pair and $M_T=\sqrt{(E_{T1}+E_{T2})^2-(p_{x1}+p_{x2})^2-(p_{y1}+p_{y2})^2}$. This variable has been studied by \cite{Edelhauser:2015ksa}.

\item The angular variable $\cos(\theta^{*})$~\cite{Barr:2005dz} where $\theta^*$ is defined as the boost invariant $\tan(\theta^*)=\tanh\left(\frac{\eta_{\ell^+}-\eta_{\ell^-}}{2}\right)$. In the case of sparticles production and decay to short chains, as slepton to leptons + $\met$~\cite{Barr:2005dz}, or sbottoms to bottom jets + $\met$~\cite{Alves:2007xt}, this variable is shown to be correlated to the production angle of the sparticles.

\item The contransverse mass $MT_c=\sqrt{2\left(\vec{p}_{T_\ell}\cdot\vec{p}_{T_\ell}+p_{T_\ell}p_{T_\ell}\right)}$ \cite{Tovey:2008ui}.

\item The angular variable $\cos(\theta_M)$~\cite{Melia:2011cu},  the cosine of the angle between
the leptons' direction of motion and the beam-axis in the centre of mass frame of the leptons. Compared to $\cos(\theta^{*})$, $\cos(\theta_M)$ is not sensitive to end point effects observed as $|\cos(\theta^{*})|\sim 1$, which diminishes the discerning power
of the $\theta^*$ variable in this region.

\end{itemize}

\begin{figure}[!ht]
  \centering
  {\includegraphics[width=.4\textwidth]{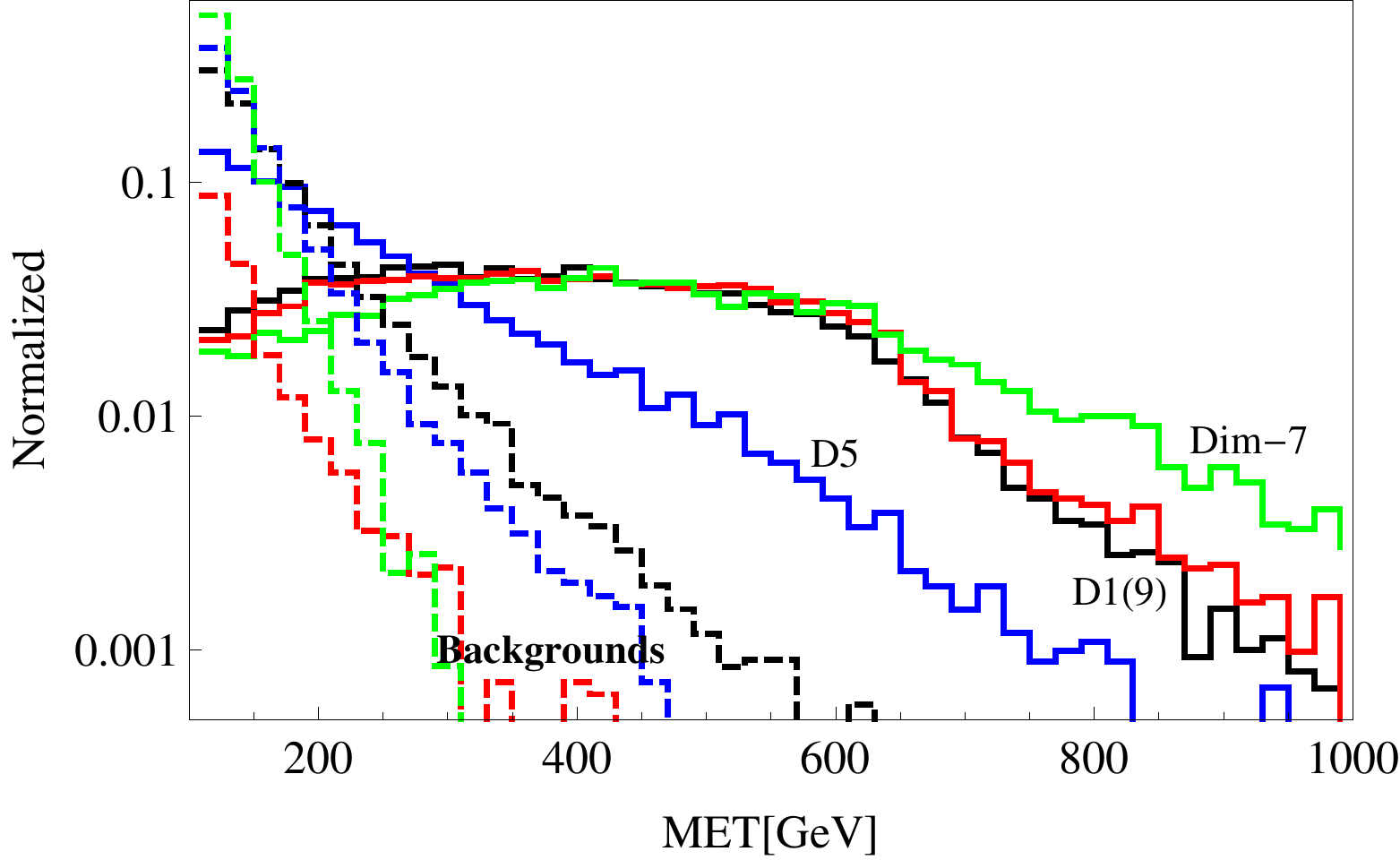}}\quad
  {\includegraphics[width=.4\textwidth]{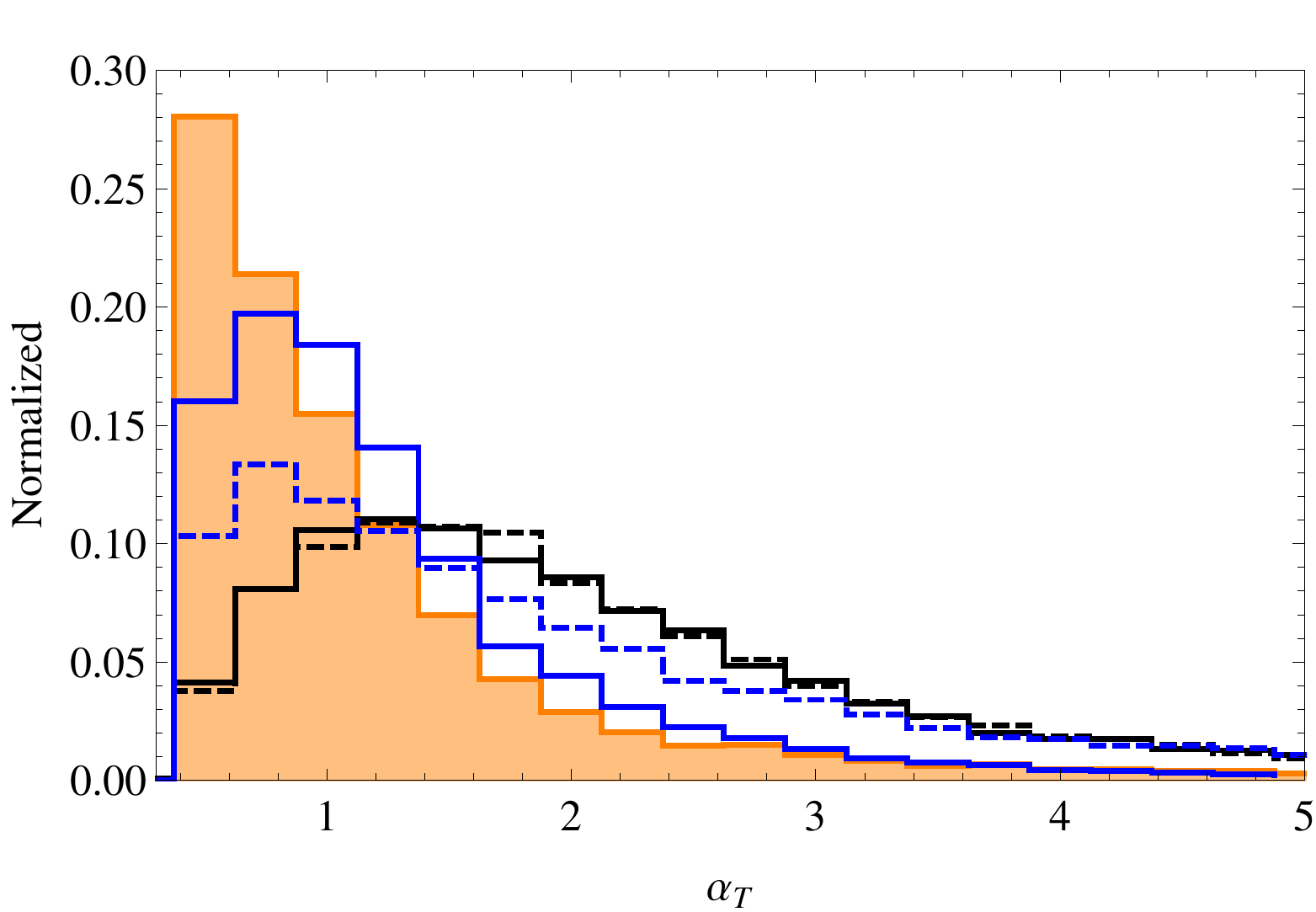}}\\
  {\includegraphics[width=.4\textwidth]{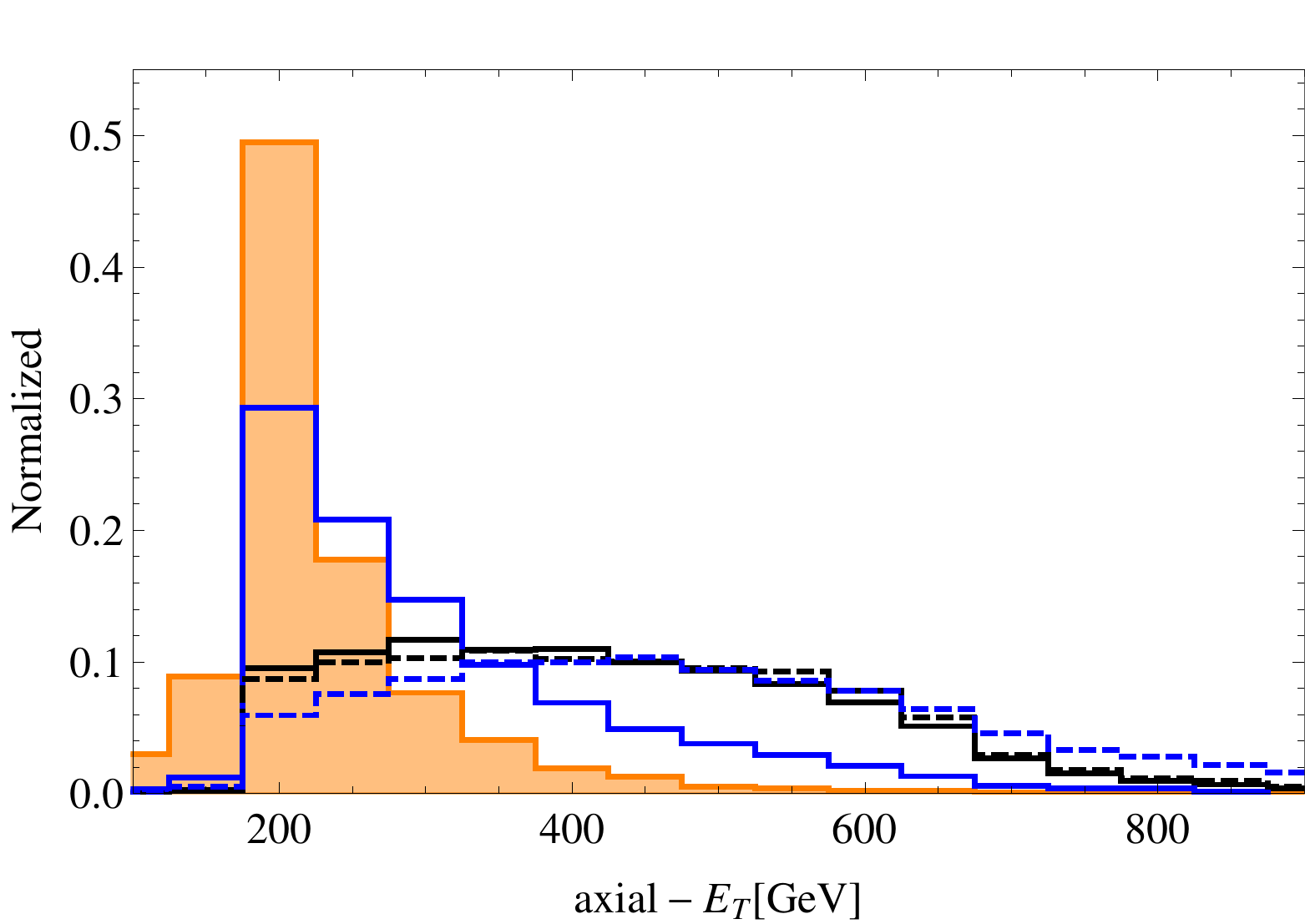}}\quad
  {\includegraphics[width=.4\textwidth]{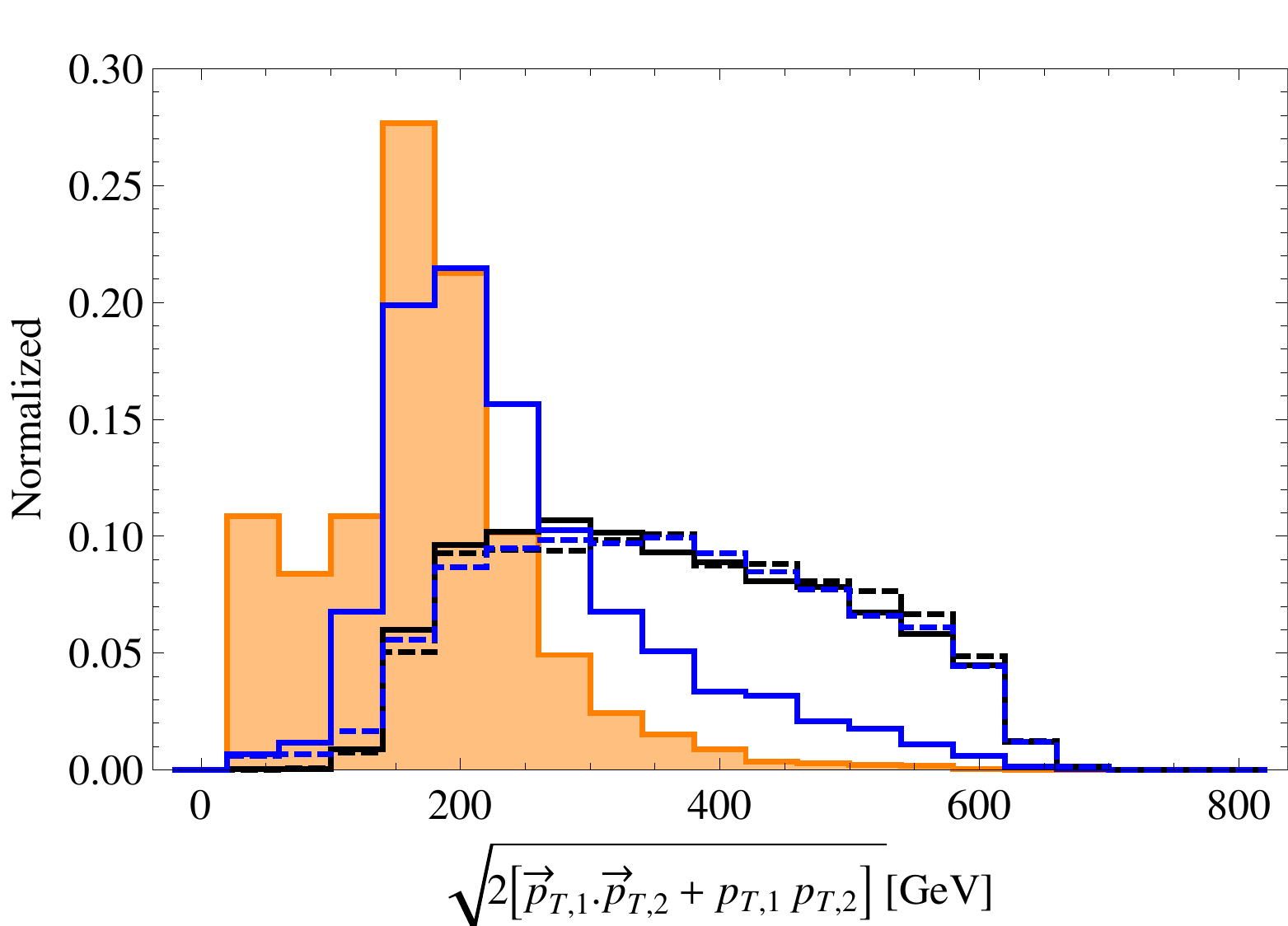}}
  \caption{Upper panel: Plot of $ \,\, \met$ (left) and $ \,\, \alpha_T$ (right). Lower panel: Plot of $ \,\, \met \, \times \, \cos\left(\Delta\phi(\vec{p}_T^{miss},\vec{p}_T^Z)\right)$ (left), and $ \,\, MT_c$ (right). Legend - upper left plot: Operator D1 (black solid line), operator D5 (blue solid line), operator D9 (red solid line), operator $ZZ \chi \chi $-dim 7 (green solid line). SM backgrounds are dotted, with $ZZ \rightarrow \ell \ell \nu \nu$ denoted by the black dotted line. Other plots: Operator D1 (black solid line), operator D5 (blue solid line), operator D9 (dotted black line), operator $ZZ \chi \chi $-dim 7 (dotted blue line), SM background (shaded orange).}  
  \label{fig:sub2}
\end{figure}

We now turn to a discussion of the normalized distributions of the above variables. In the upper left panel of Fig.~(\ref{fig:sub2}) we show the $\met$ distribution for the SM backgrounds (dashed lines) and EFT interactions (solid lines). The black, blue, red, and green solid lines show the distributions for $\tilde{\hbox{D}1}$, D5, D9, and the dim-7 operators. Not surprisingly, vector (D5) and vector-axial couplings (D8) (not shown in the figure but very similar to D5) have the closest resemblance to the $ZZ$ background (black dashed). The other EFT interactions, in turn, show harder distributions. The other panels of Fig.~(\ref{fig:sub2}) display those variables which are correlated to $\met$. In the upper right panel, we show the variable $\alpha_T$, while in the bottom left and right panels we show the axial-$\met$ and contransverse mass $MT_c$, respectively. In these figures, the SM background is shaded in orange, while the solid black and blue lines denote operators $\tilde{\hbox{D}1}$ and D5, respectively, while dotted black and blue lines denote D9 and the dim-7 operators, respectively.

\begin{figure}[!ht]
  \centering
  {\includegraphics[width=.4\textwidth]{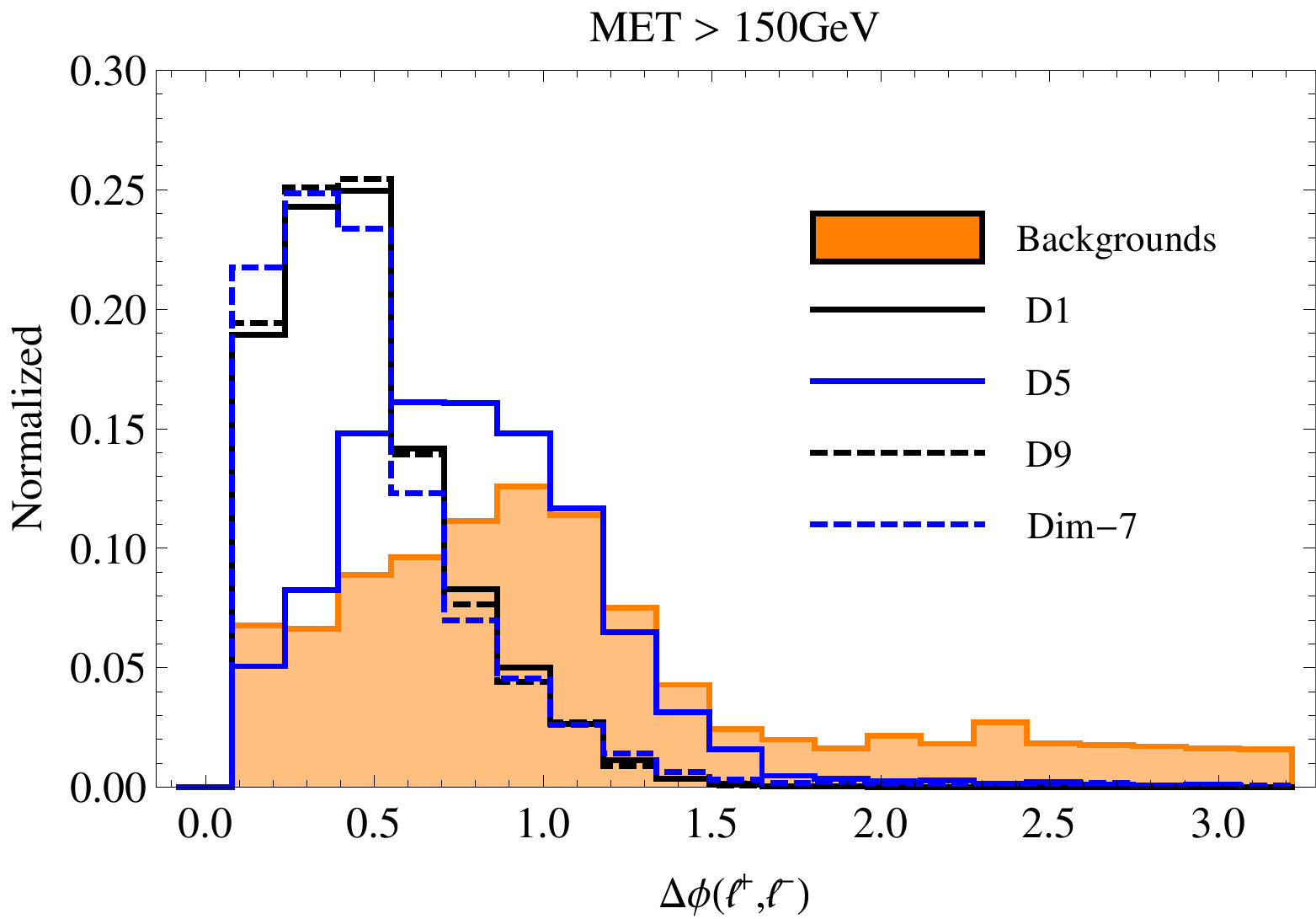}}\quad
  {\includegraphics[width=.4\textwidth]{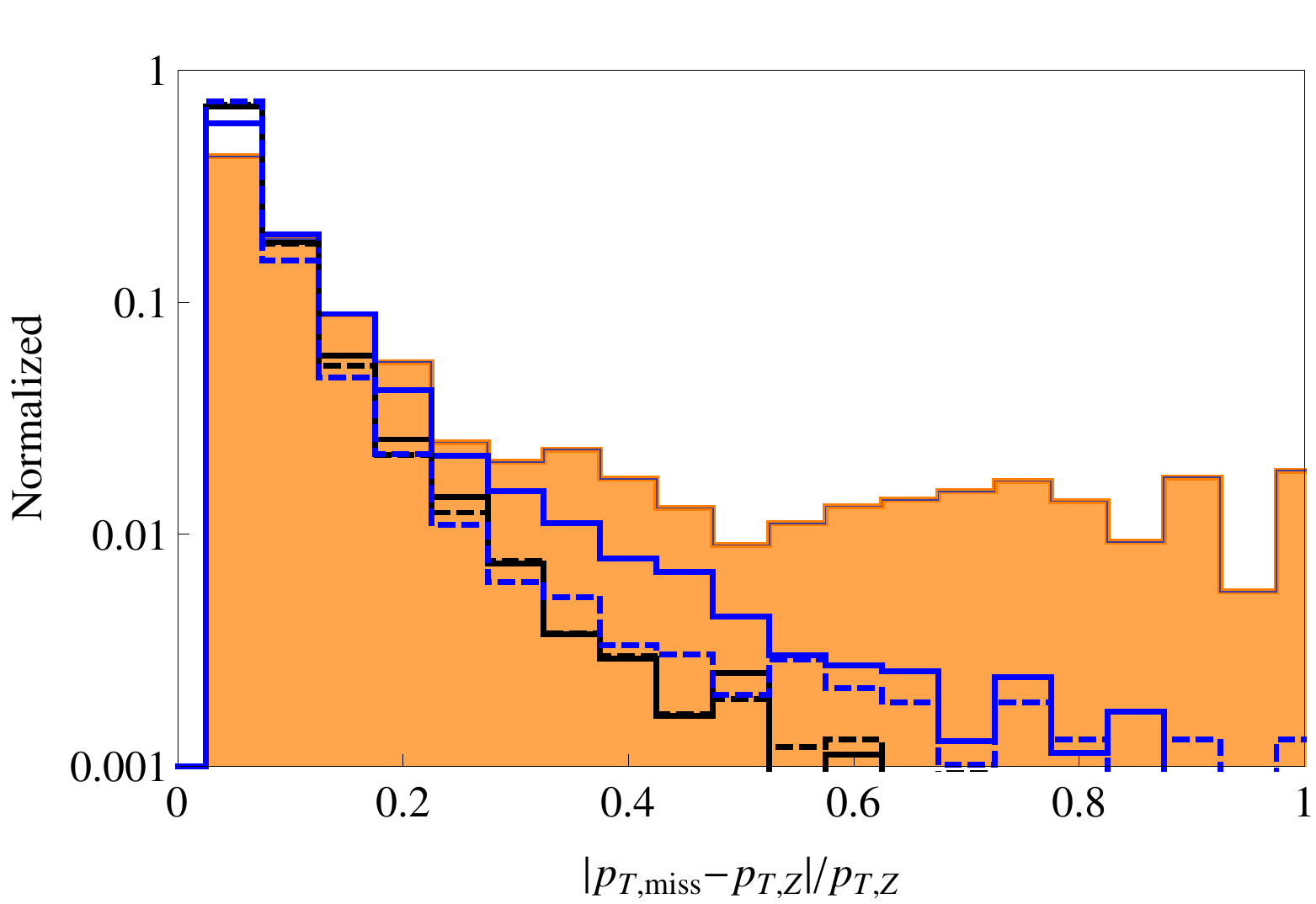}}\\
  {\includegraphics[width=.4\textwidth]{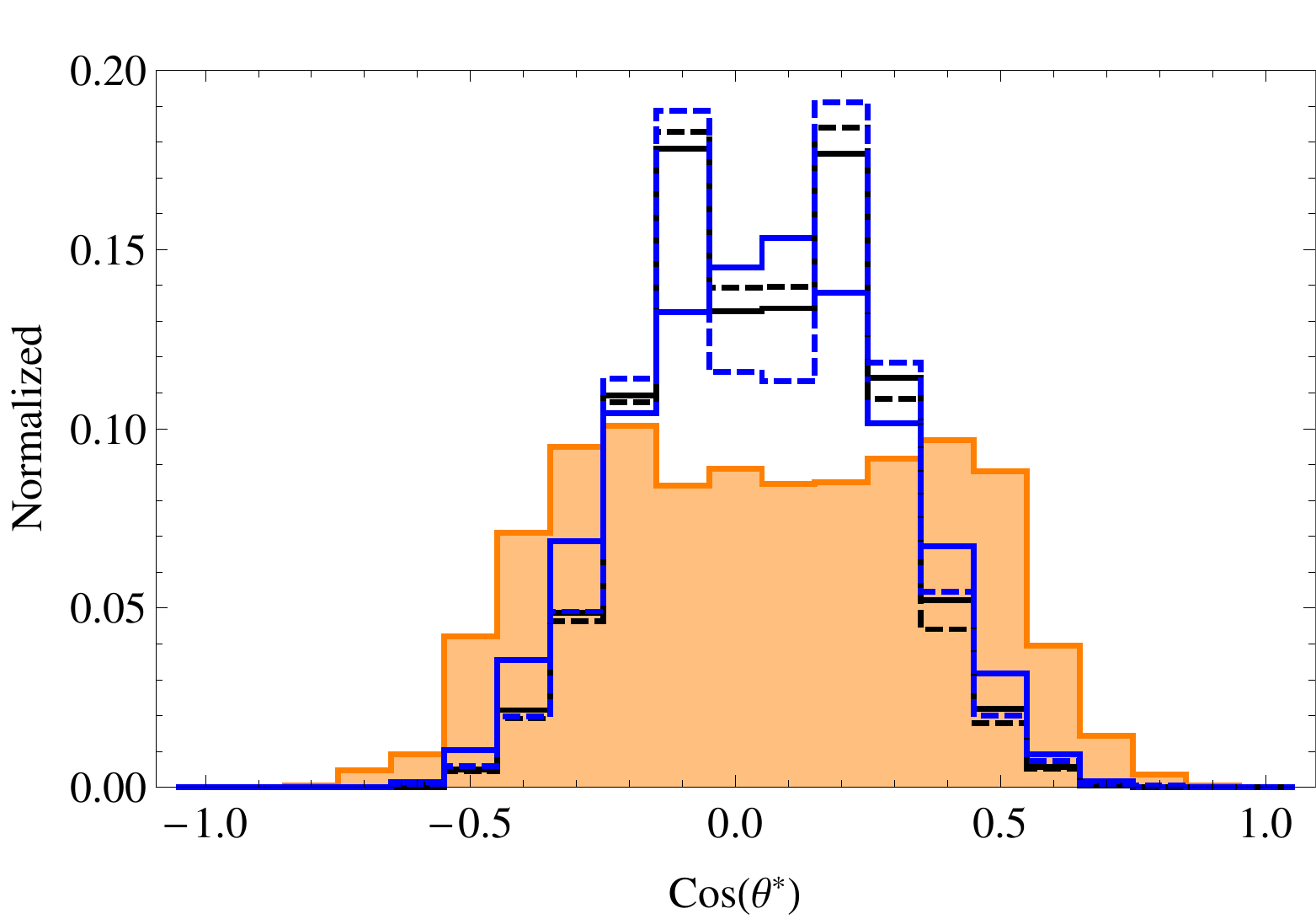}}\quad
  {\includegraphics[width=.4\textwidth]{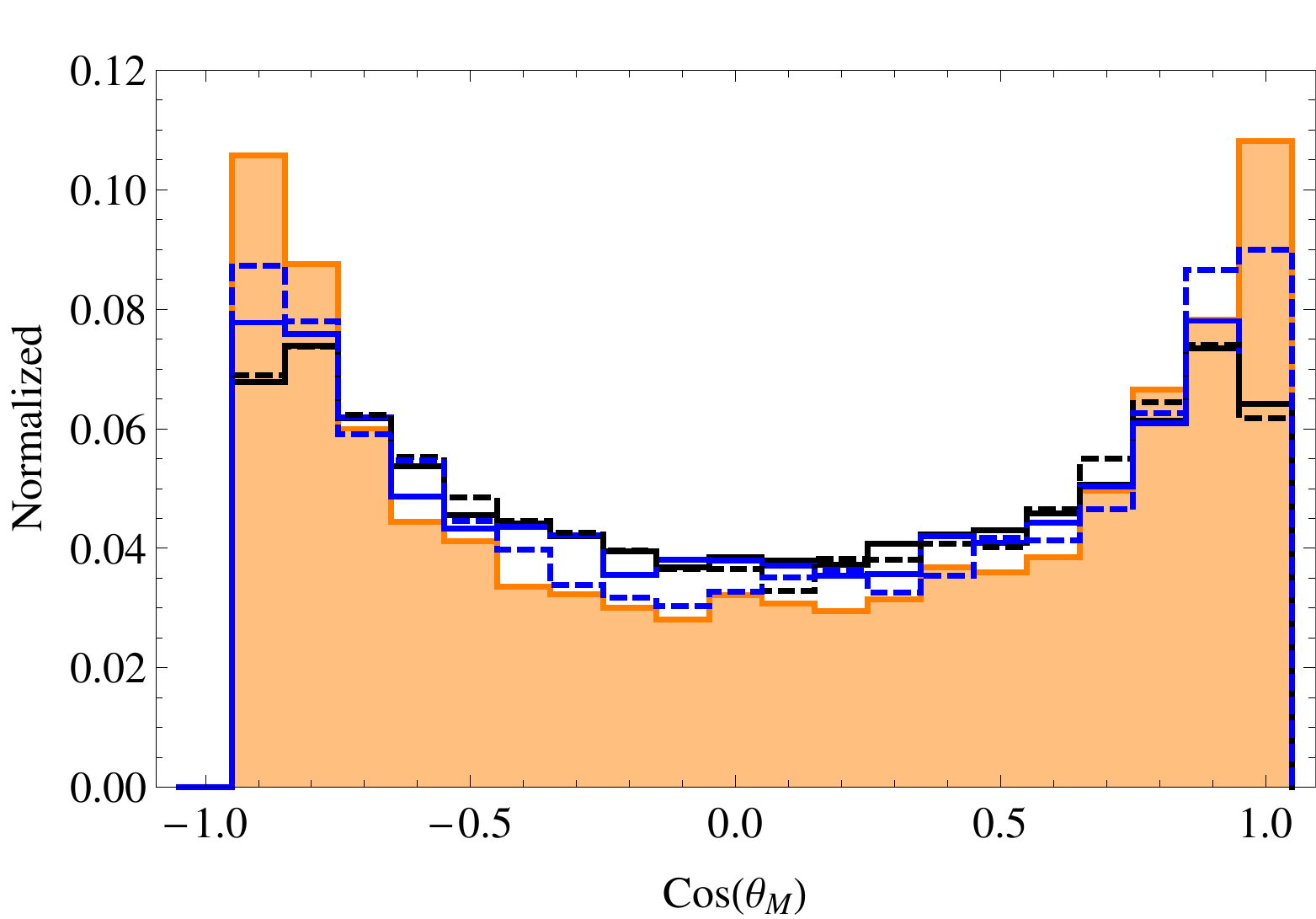}}
  \caption{Upper panel: Plot of $ \Delta\phi(\ell^+,\ell^-)$ (left) and $\,\, |\met - p_T^Z|/p_T^Z$ (right). Lower panel: $\cos(\theta^{*})$ (left) and  $\cos(\theta_M)$ (right). Legend: Operator D1 (black solid line), operator D5 (blue solid line), operator D9 (dotted black line), operator $ZZ \chi \chi $-dim 7 (dotted blue line), SM background (shaded orange).}
  \label{fig:sub1}
\end{figure}

In the Fig.~(\ref{fig:sub1}) we show the angular variables used in the construction of the multivariate discriminant after the selection cuts and a missing transverse cut $\met > 150$ GeV. In the upper left panel we show the azimuthal angle between the leptons. This is more peaked to small values for EFT operators than the SM backgrounds, a consequence of harder missing transverse momentum of dark matter which makes the signal lepton pairs more collimated. In other words, the $Z$ boson transverse momentum gets more correlated to the dark matter pair $p_T$ as can be seen at the upper right panel where, again, the EFT interactions lead to smaller fractional $p_T$ difference $|\met - p_T^Z|/p_T^Z$ as compared to the backgrounds. For the $\cos(\theta^*)$ and $\cos(\theta_M)$ distributions we see basically the same effect of enhanced collimation of EFT interactions compared to the SM events.  

As a general remark we note that the vector effective interaction is, again, the most similar to the SM backgrounds (as well as D8), while $\tilde{\hbox{D}}$1, D9 and Dim-7 are similar to each other but less similar to the backgrounds. Of course, these features will have an impact on the discerning power of the discriminants for each EFT operator.

\begin{figure}[!ht]
  \centering
  {\includegraphics[width=.32\textwidth]{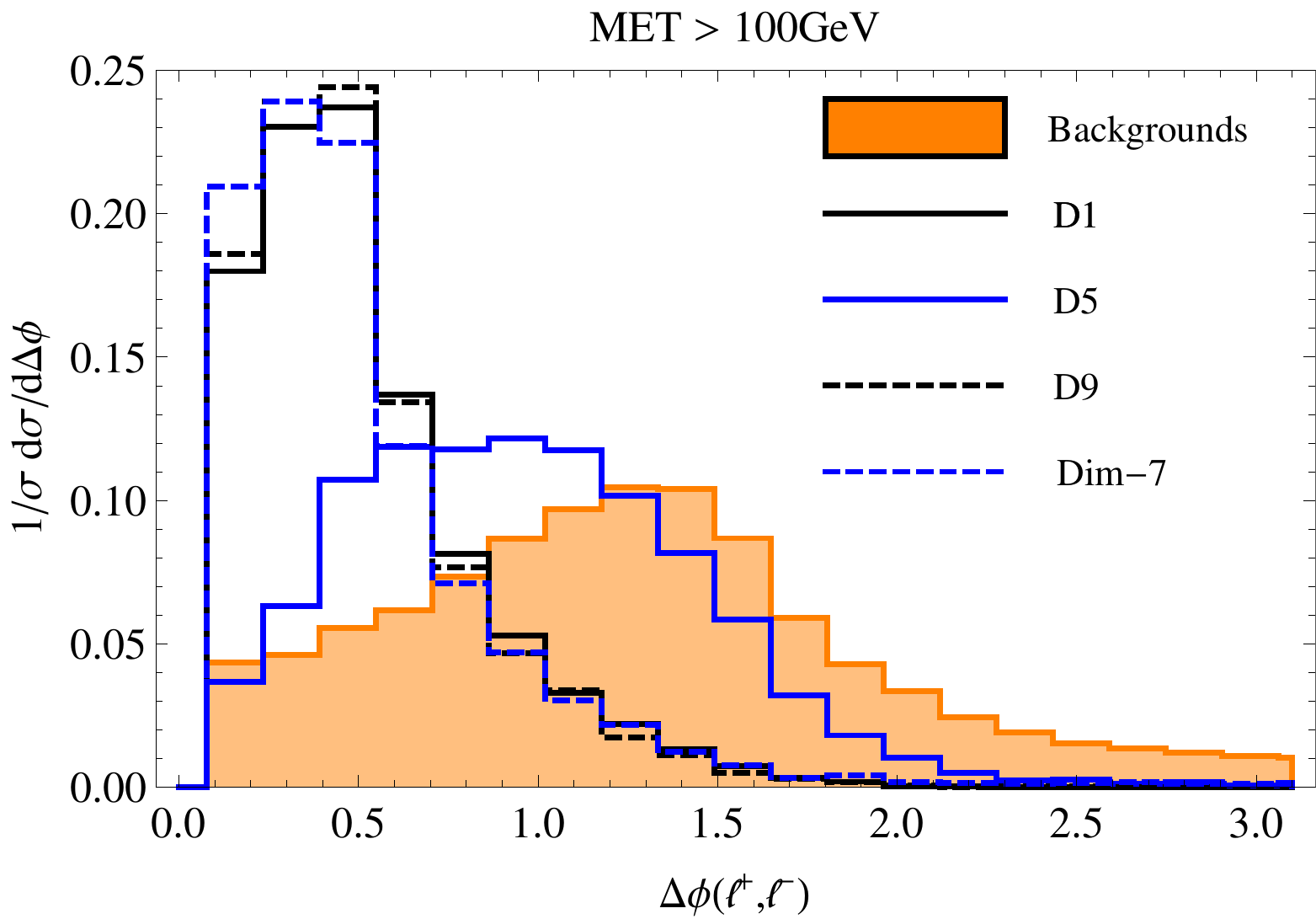}}\quad
  {\includegraphics[width=.32\textwidth]{dist1_150.pdf}}\quad
  {\includegraphics[width=.32\textwidth]{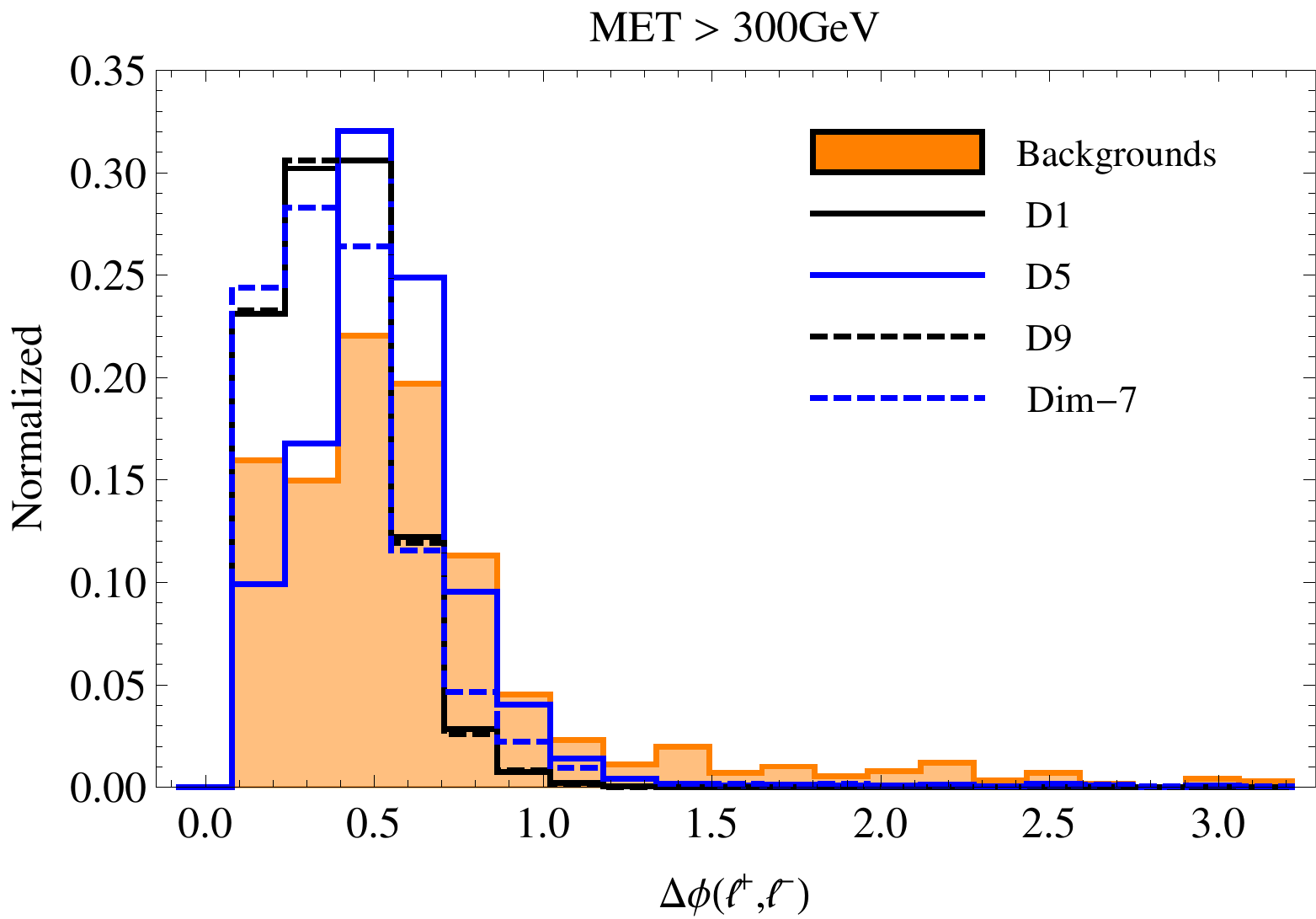}}\quad
  \caption{From left to right, the impact of increasing the $\met$ cut ($\met > 100, \,\, 150, \,\,$ and 300 GeV, respectively) on the $\Delta\phi(\ell^+,\ell^-)$ variable.}
  \label{fig:sub3}
\end{figure}

Demanding harder $\met$ cuts to further separate signal events from backgrounds has a deleterious effect on the discrimination power of the shapes of these distributions. As $\met$ increases, the $Z$ momentum gets more correlated to the momentum of the dark matter pair, and higher correlation translates into increased similarity between signal and backgrounds distributions. This can be seen in Fig.~(\ref{fig:sub3}) where we show, from left to right, the impact of increasing the $\met$ cut on the $\Delta\phi(\ell^+,\ell^-)$ variable.

\section{Systematic Uncertainties and Statistical Significance} \label{systuncert}

Systematic uncertainties frequently limit the reach of experiments, mainly in processes with small signal to background ratio. Multivariate techniques can be used to overcome these limits, extracting more information and circumventing the constraints imposed by systematics in the statistical significance of a signal. A very good example is the single top measurement first observed by the CDF collaboration~\cite{Aaltonen:2010jr} where the combination of four multivariate techniques was necessary to extract the signal in spite of the small signal to background ratio and a large number of sources of systematic uncertainties.

Incorporating systematic uncertainties in the computation of the statistical significance level of a signal can be done in many ways and the subject itself has been debated in the statistical community~\cite{Cousins}. 

There are many sources of systematics in a real experiment and taking all of them into account in a simulation is difficult. However, in a discovery analysis, the loss of statistical significance is, roughly, an effect of the widening of the probability density distribution of the chosen test statistic associated to the background hypothesis due to the presence of the systematic uncertainties. This can be more easily understood in terms of the naive significance metric given by
\be
Z_{sb}=\frac{S}{\sqrt{B+(\varepsilon_{sys}B)^2}}
\label{zsb}
\ee
where $S$ is the number of signal events, $B$ the number of backgrounds events, and $\varepsilon_{sys}B$ the systematic uncertainty on the backgrounds in the limit of high statistics i.e. $B\gg 1$. 

It is easy to invert this relation to obtain the minimum integrated luminosity required to reach a given statistical significance $Z_{sb}$. If $\sigma_S$ and $\sigma_B$ are the signal and backgrounds cross sections, respectively, it can be shown that $\sigma_S/\sigma_B > Z_{sb}\varepsilon_{sys}$ is a constraint on the achievable significance. If systematics are too large, a $5\sigma$ discovery, for example, is not possible no matter how much data has been accumulated.

The formula~(\ref{zsb}) given above takes into account only the total number of signal and background events and increasing the signal to background ratio can be done only in a cut-and-count analysis. This is our approach in this work, constructing a distribution where signal and backgrounds events are well separated and imposing a cut to clean up the signal events more efficiently than the $\met$ distribution. 

Systematic uncertainties in the shape of the distributions used in the multivariate analysis might also be important, but that is beyond the scope of our work. We note that the results obtained in our more simple approach to systematics are sufficient to show that taking these errors into account is crucial to reliably estimate the reach of the experiment.

For the purpose of calculating the signal significance in the presence of systematic uncertainties in the number of backgrounds events we use, instead of Eq.~(\ref{zsb}) which is shown to overestimate the significance for $\varepsilon_{sys}<1$~\cite{Cousins}, the metric proposed in~\cite{LiMa}, $Z_{PL}$, which is simple to use and is shown to be an excellent approximation to the consistent frequentist computation using Poisson distributions. Moreover, we found that among the three more reliable methods to compute the statistical significance incorporating systematic uncertainties, $Z_{PL}$ is the most conservative  one. More details on these metrics are given in Appendix A. 



%
%

%
%

 The background estimation and signal extraction strategies would depend on upgraded detector designs and trigger conditions, especially in high pileup conditions. However, based on current data, we can give a very simple estimation of the systematic uncertainties on the $ZZ \rightarrow \ell \ell \nu \nu$ background. In \cite{Aad:2012awa}, the ATLAS collaboration studied the process $ZZ \rightarrow \ell \ell \nu \nu$ at 7 TeV with 4.6 fb$^{-1}$ of data, with an event count of $39.3 \pm 0.4 \pm 3.7$, where the first and second uncertainties are statistical and systematic, respectively. The systematic uncertainties include contributions from uncertainties in the lepton reconstruction efficiency and the lepton energy scale and resolution, the missing transverse energy modeling, the jet veto uncertainty and uncertainties in the trigger efficiency, PDF and scale, and generator modeling and parton shower. In a recent study from the CMS collaboration, \cite{Khachatryan:2015pba}, the cross section for the process $pp \rightarrow ZZ \rightarrow \ell \ell \nu \nu$ was given as $88^{+11}_{-10} (stat)^{+24}_{-18} (syst)$ fb at 8 TeV with $Z$ bosons in the mass range 60 to 120 GeV. From the above estimates, it is clear that systematic uncertainties $\sim \mathcal{O}(10\%)$ are possible.

 The other SM background that is important besides $ZZ + j$ includes contributions from $t \overline{t}, WW, WZ$. However, as we are going to show, after applying the MVA and jet veto selections they become negligible and the only relevant sources of systematic errors are those associated to the $ZZ$ backgrounds as we described here.

\section{Multivariate Analysis and Results} \label{MVA}

One robust method for combining several sensitive variables is the multivariate likelihood function~\cite{Aaltonen:2010jr,Abazov:2008kt,Barlow,Gomez:2014lva}, a Bayesian learning classifier~\cite{Prosper:2008zz}. The relative probabilities of finding an event in histograms of each input variable described above is used, comparing between signal and background. 

Binned probability density functions for each input variable are used to construct the likelihood function $\mathcal{D}_k$, where $k$ denotes the event class ($k = 1$ for signal, $k=2,3,4,5$ for $ZZ$, $WW$, $ZW$ and $t\overline{t}$ backgrounds, respectively). We denote by $f_{ijk}$ the probability that an event from the sample $k$ will populate the bin $j$ of the kinematic variable $i$. The probabilities are normalized to
\be
\sum_{j}f_{ijk} \, = \, 1 \,\,
\ee
for every variable $i$ and event sample $k$. The likelihood function is computed as follows. For each reconstructed variable $i$, the bin $j$ in which the event falls is obtained. The quantities 
\be
p_{ik} \, = \, \frac{f_{ijk}}{\sum_{m=1}^5 f_{ijm}}
\ee
are computed for every variable $i$ and all values of $k$. These quantities $p_{ik}$ are finally used to calculate the likelihood function
\be
\mathcal{D}_k \, = \, \frac{\prod_{i=1}^{n_{var}} p_{ik}}{\sum_{m=1}^5 \prod_{i=1}^{n_{var}} p_{im}}
\ee
where $n_{var}$ is the number of input variables. The quantity $\mathcal{D}_{LF}\equiv \mathcal{D}_1$, corresponding to the signal likelihood function, is referred to as the likelihood discriminant. 
\begin{figure}[!ht]
  \centering
  {\includegraphics[width=.48\textwidth]{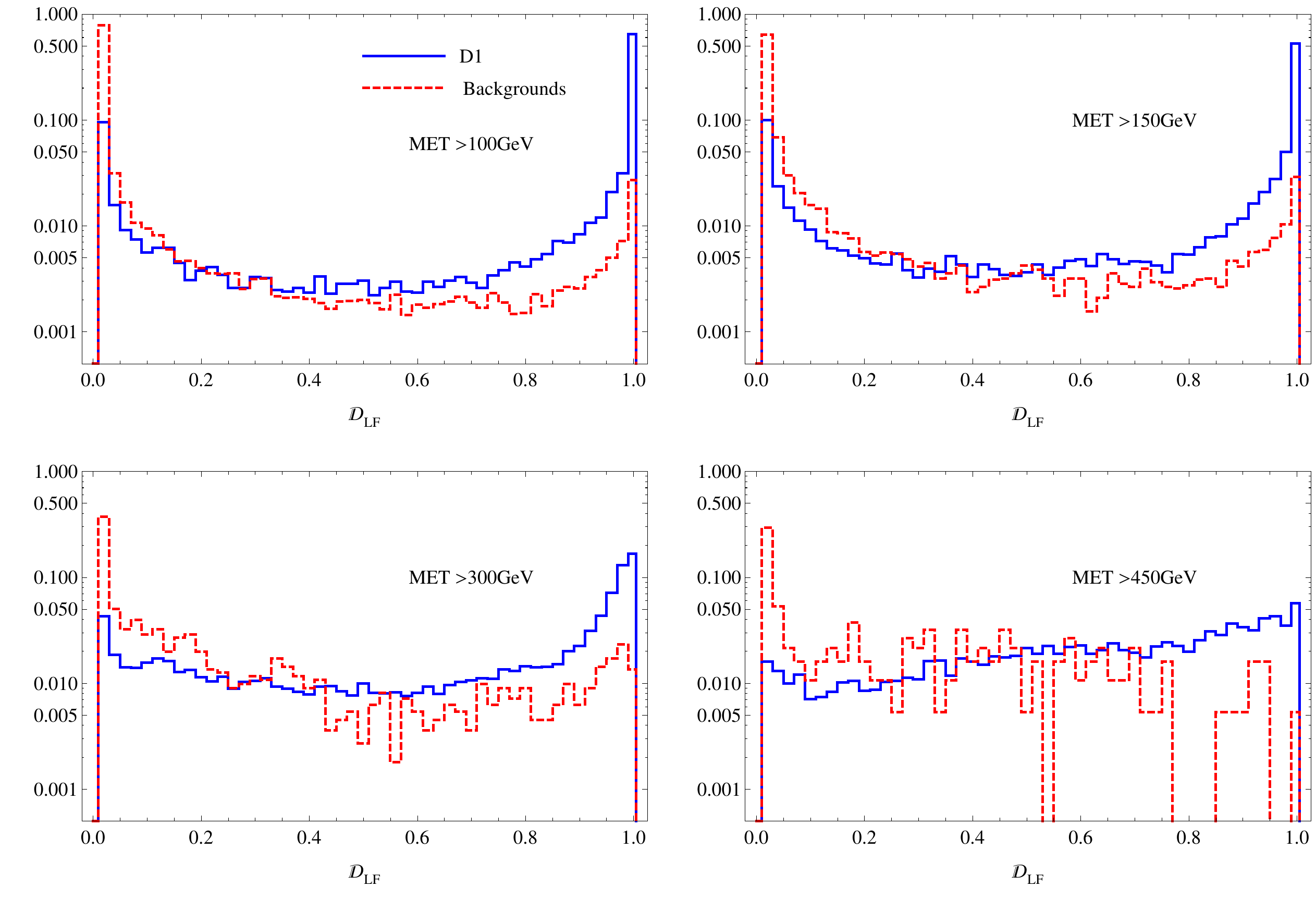}}\quad
  {\includegraphics[width=.48\textwidth]{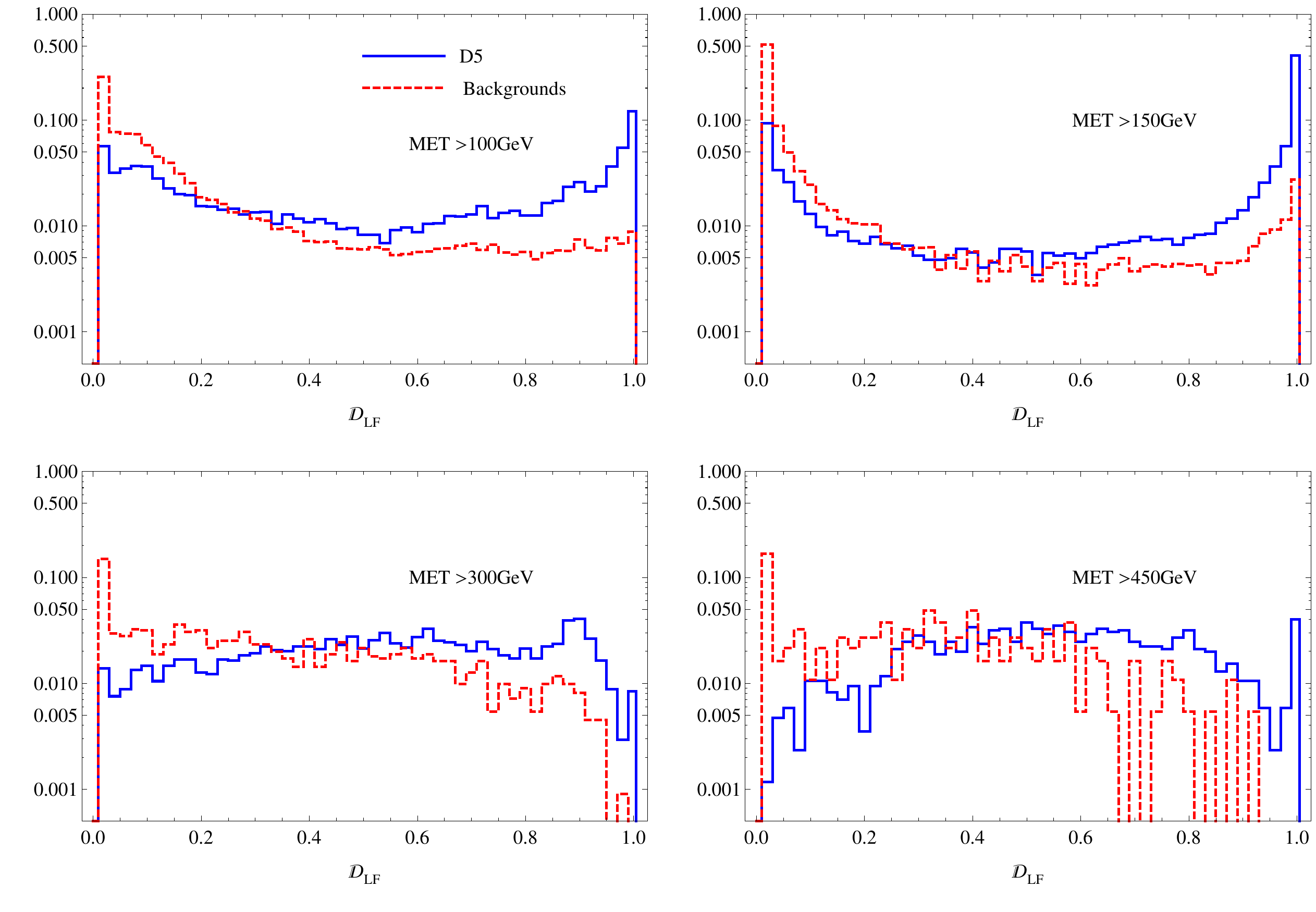}}\\
  {\includegraphics[width=.48\textwidth]{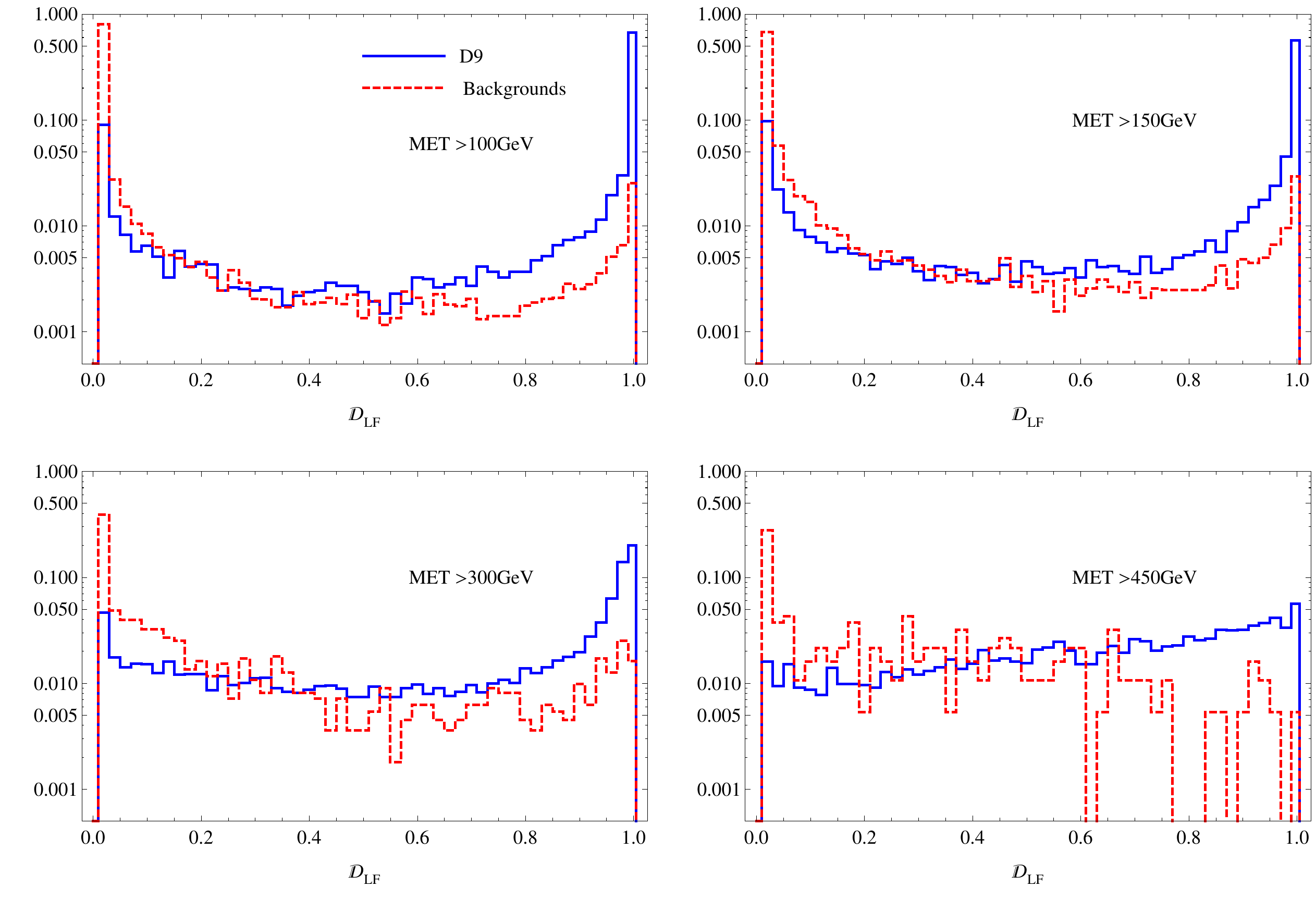}}\quad
  {\includegraphics[width=.48\textwidth]{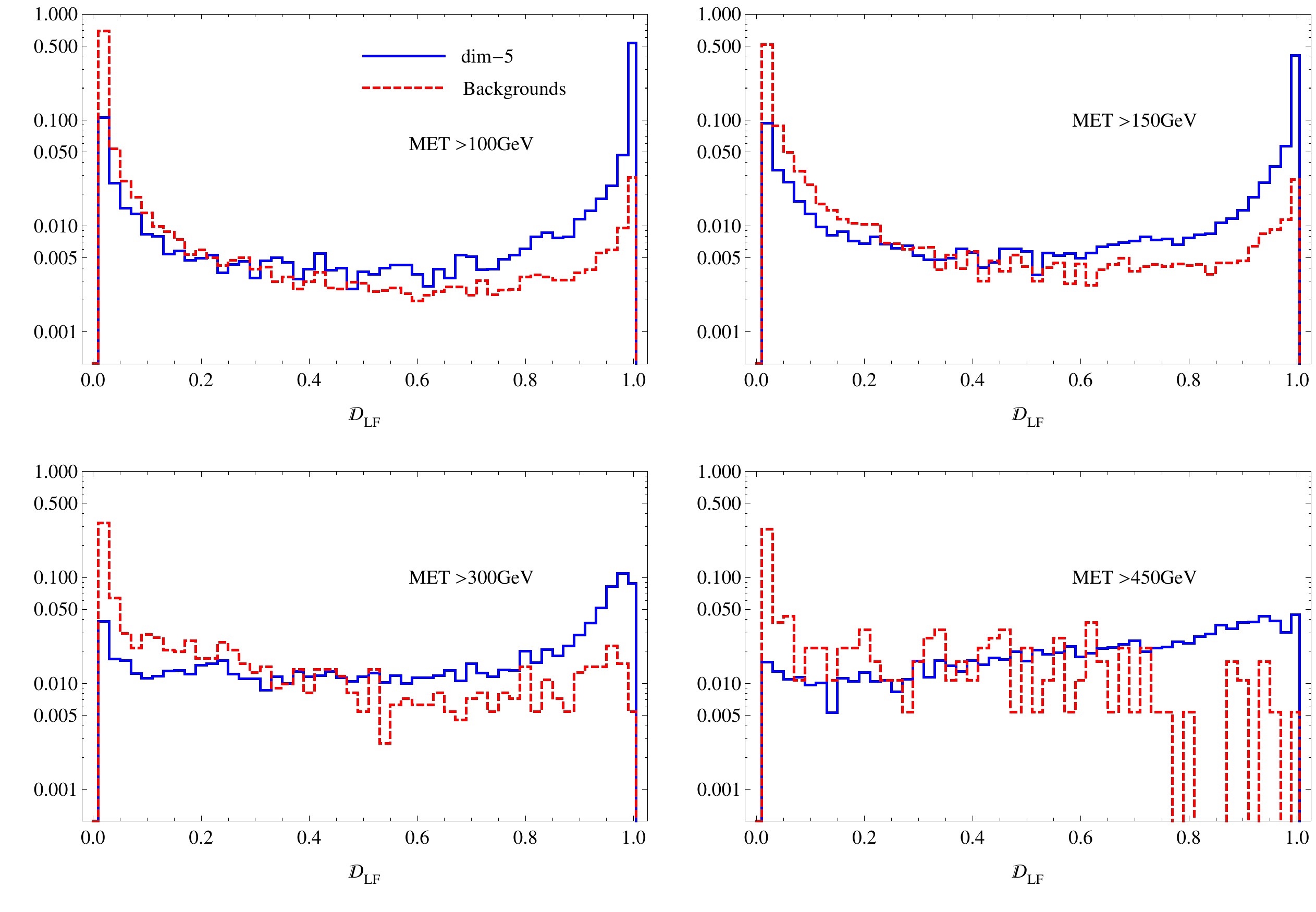}}\\
  {\includegraphics[width=.48\textwidth]{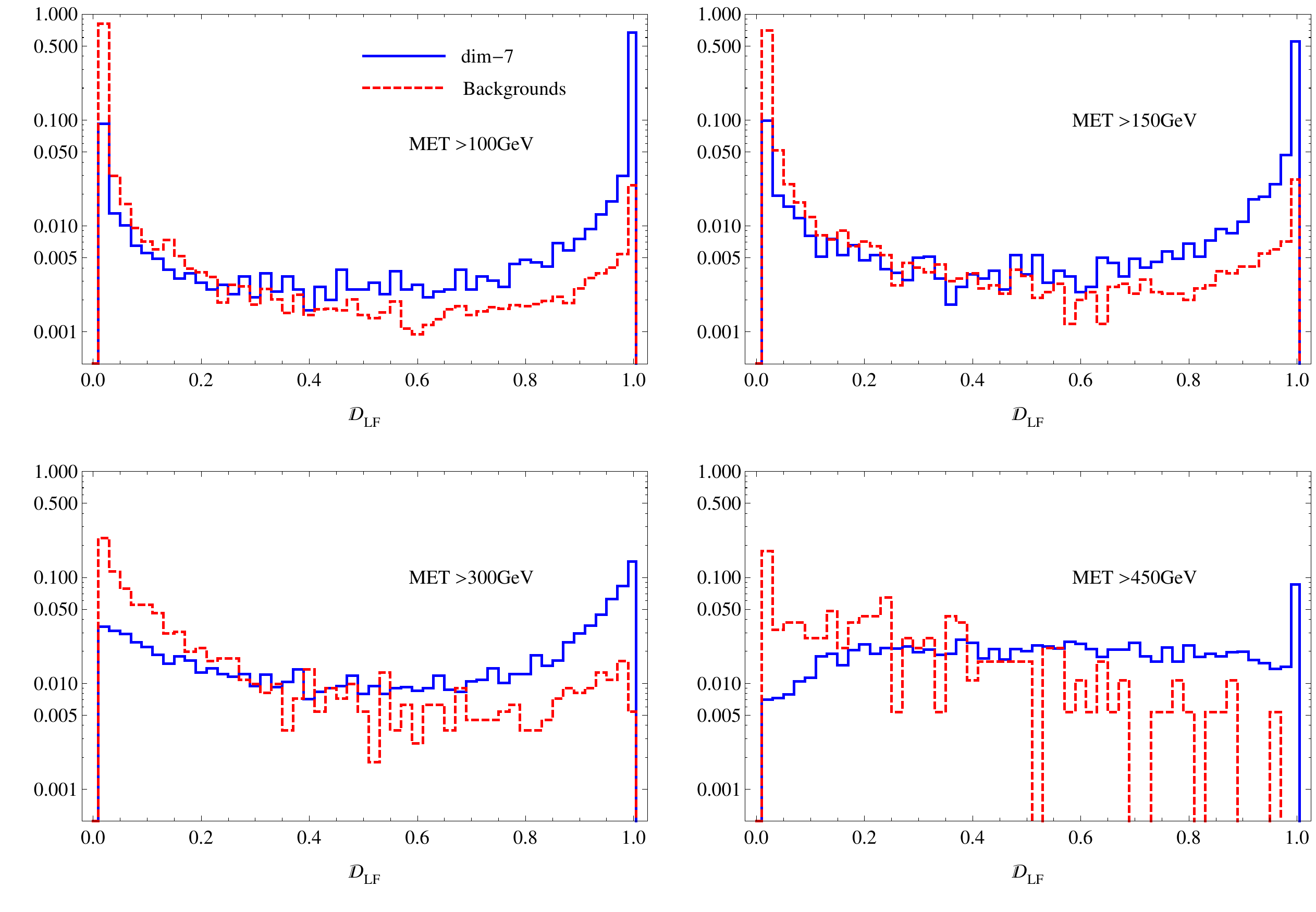}}\quad
  {\includegraphics[width=.48\textwidth]{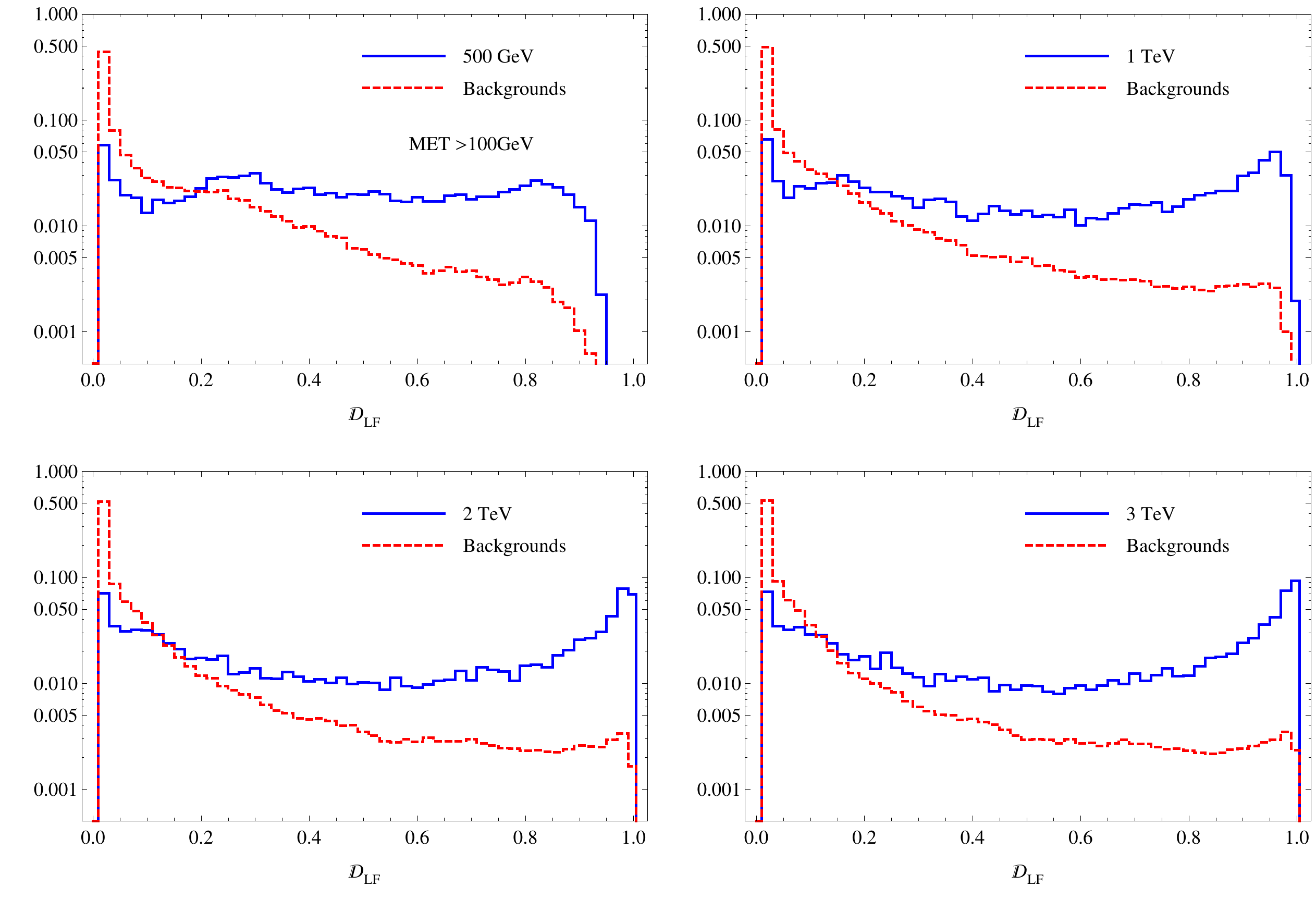}}
  \caption{The likelihood function $\mathcal{D}_{LF}$ is displayed for operators $\tilde{\hbox{D}}1$, D5, D9, dim-5, dim-7, and the $\zp$ model. The blue line shows the distribution for signal, while the red line shows the distribution for background. The EFT plots show the effect of the $\met$ cut on the discerning power of the likelihood discriminant, while the $\zp$ plots, at the lower right panels, show that heavier $\zp$ are easier to separate from the SM backgrounds. For $\zp$ we fixed $\met > 100$ GeV.}
  \label{fig:discrim}
\end{figure}

We note that the likelihood function does not use correlations between the variables, although since the distributions are obtained from the fully simulated Monte Carlo data, any correlations are included in the modeling. More sophisticated analyses take advantage of the correlations between the variables.

In Fig.~(\ref{fig:discrim}), we display the  likelihood function $\mathcal{D}_{LF}$ distributions for operators $\tilde{\hbox{D}}1$, D5 (D8 is very similar to D5), D9, dim-5, and dim-7 after cuts on $\met$ have been applied. The blue line shows the distribution for signal, while the red line shows the distribution for background. It is clear that the signal peaks near $\mathcal{D}_{LF} \, = \, 1$, while the background peaks near $\mathcal{D}_{LF} \, = \, 0$. Nevertheless, we observe smaller signal peaks in the background region, where the likelihood discriminant classified signal events as background events, and {\it vice-versa}. This can be understood if one keeps in mind that leptons from both the signal and the dominant background $ZZ$ are the yields of a $Z$ boson decay, and thus it is natural that a fraction of signal events are misidentified as background events. The most distinctive features of the kinematic variables are driven by the dark matter interactions and the identity of the source of missing energy - the neutrinos in the SM case, and the $\chi$ DM in the case of new physics parametrized by the EFT operators.

A harder $\met$ cut results in EFT signal and background distributions for the likelihood function that look similar. This happens due to the fact that the signal and background shapes of the kinematic distributions start to become similar as the $\met$ cut is increased as we have shown in Fig.~(\ref{fig:sub3}). To obtain optimal reach, one thus has to balance between the $\met$ cut and the cut on $\mathcal{D}_{LF}$. The $\met$ cut is chosen to be $\met \, > 100,\, 150,\, 300$ and $450$ GeV. It is clear that increasing the $\met$ cut gradually makes the signal and background distributions of $\mathcal{D}_{LF}$ similar, reducing the efficiency of the $\mathcal{D}_{LF}$ cut. The same behavior concerning $\met$ cuts were observed in the case of $\zp$ events.

At the lower right panels of Fig.~(\ref{fig:discrim}) we show the likelihood distributions for the $\zp$ model for $M_\zp=0.5,\, 1,\, 2$ and 3 TeV fixing $\met > 100$ GeV. In this case, $\mathcal{D}_{LF}$ performs better for heavier $\zp$. This is expected, since a lighter $\zp$ tends to look like a SM $Z$ boson, that is, the $Z\zp$ process become less distinguishable from the $ZZ$, in spite of the fact that $\met$ is due a heavy DM in the signal case. Compared to EFT operators, the $\zp$ discriminant distributions discern better signal from backgrounds. 

%
%

\begin{figure}[!ht]
  \centering
{\includegraphics[width=.45\textwidth]{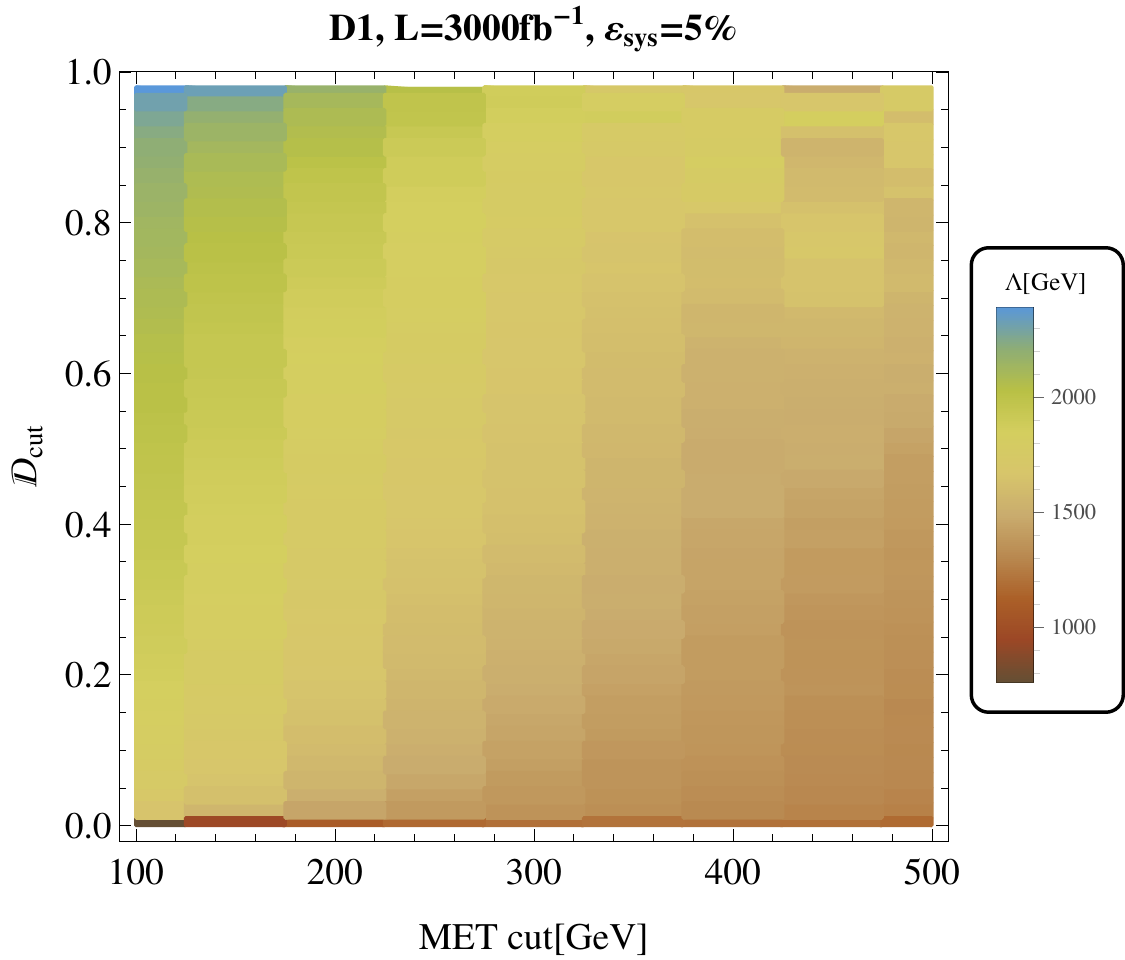}}\quad
{\includegraphics[width=.45\textwidth]{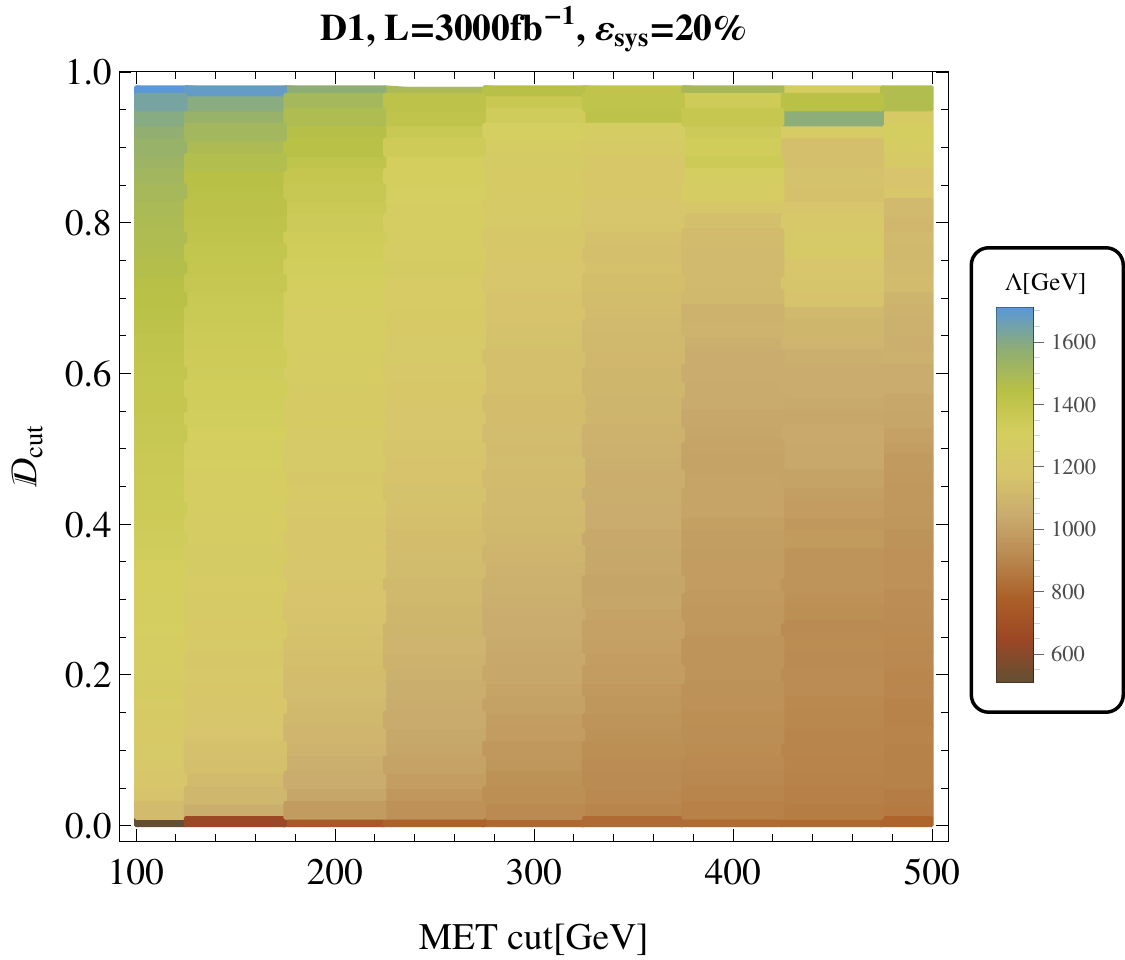}}\\
{\includegraphics[width=.45\textwidth]{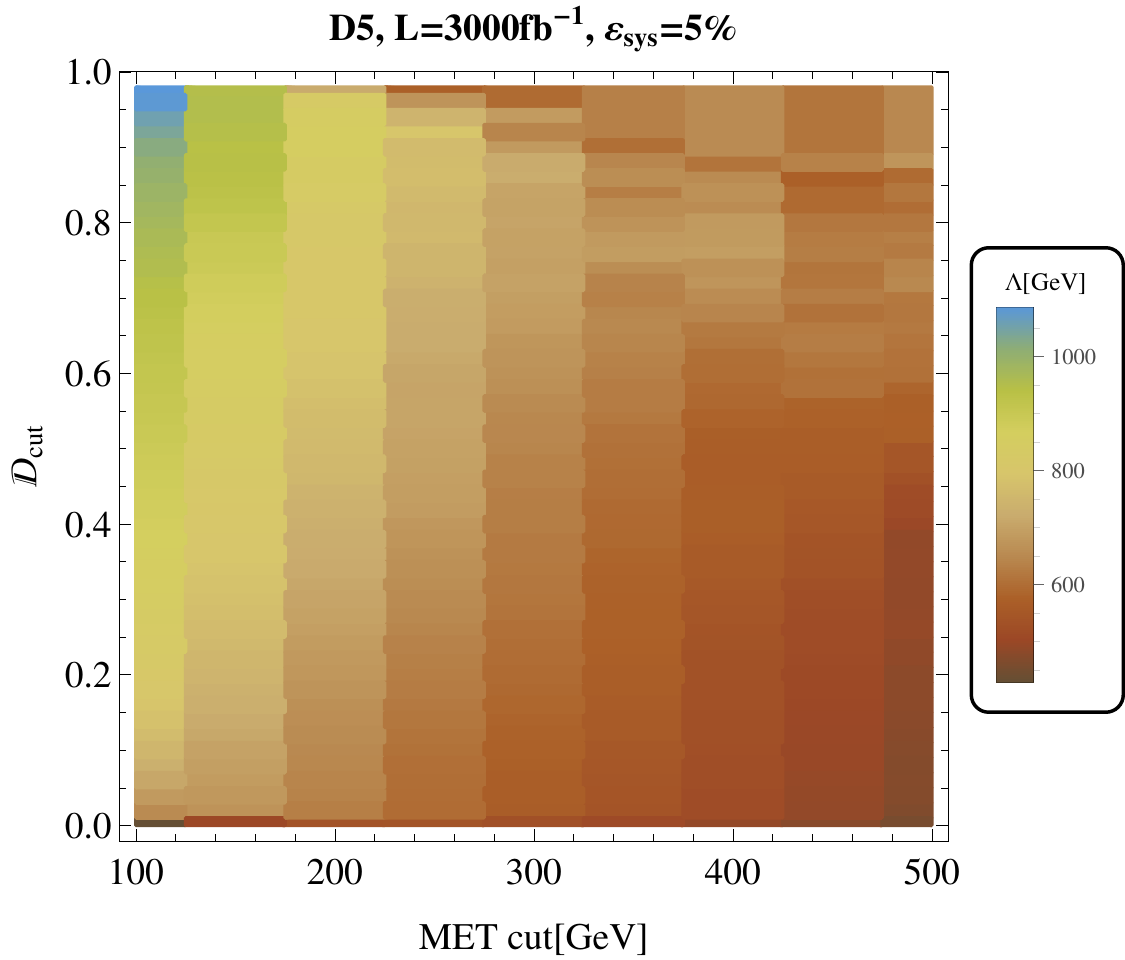}}\quad           
{\includegraphics[width=.45\textwidth]{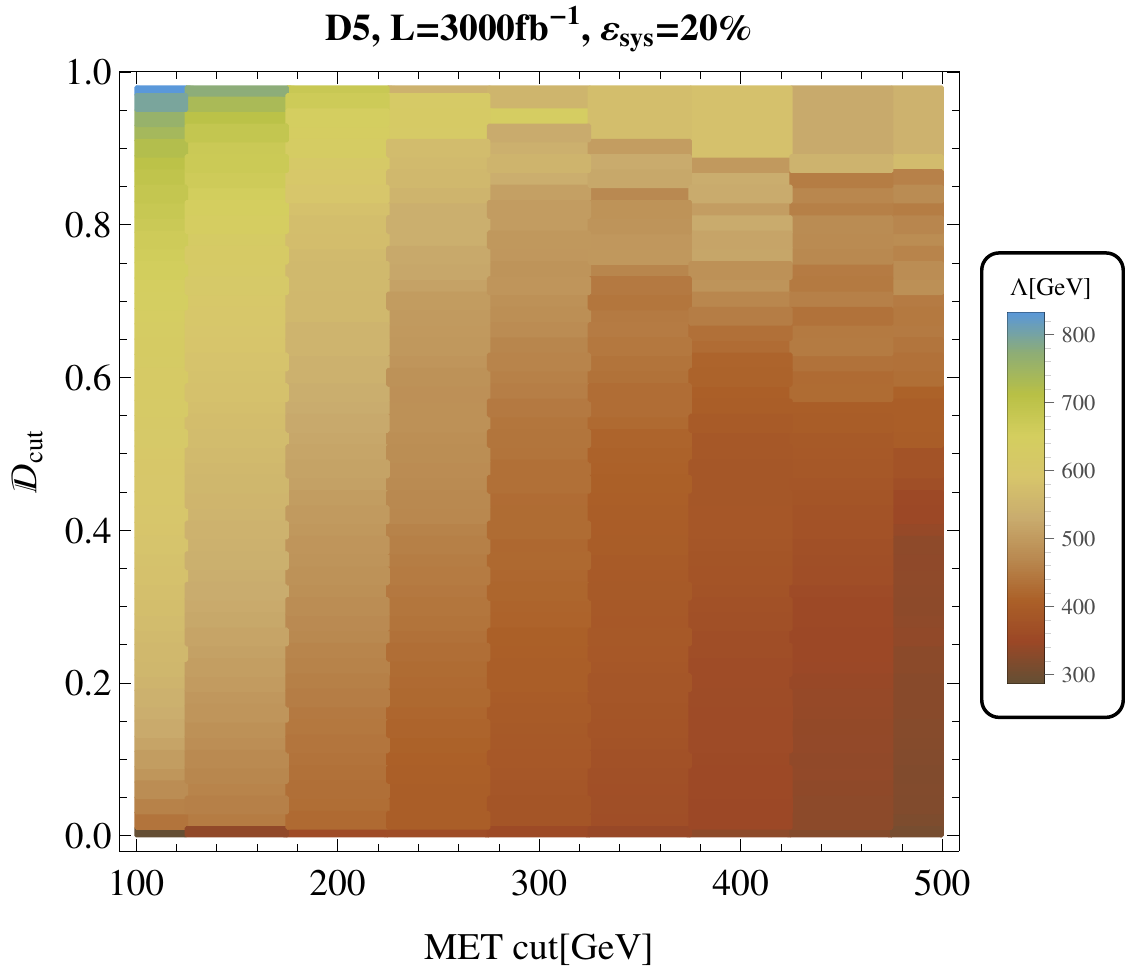}}\\
{\includegraphics[width=.45\textwidth]{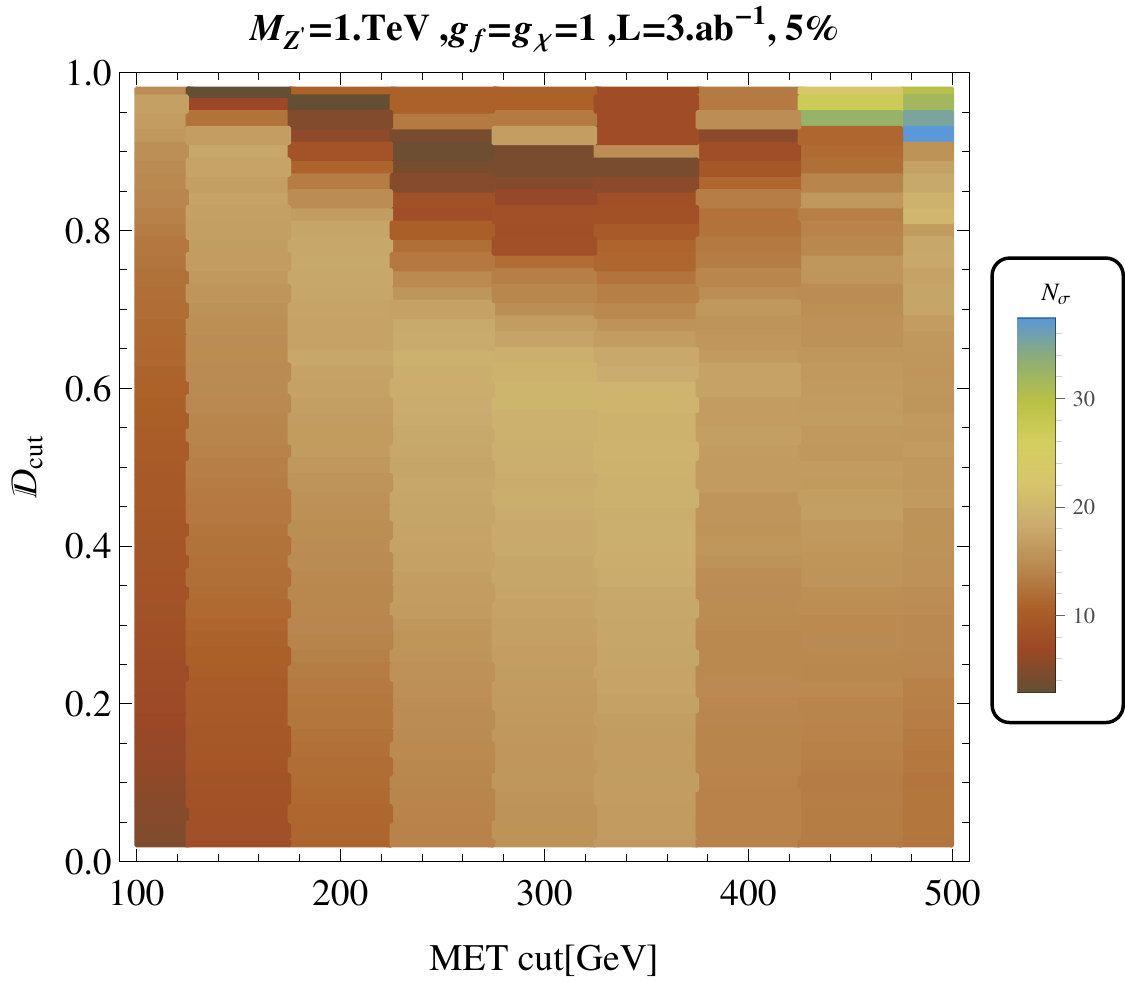}}\quad
{\includegraphics[width=.45\textwidth]{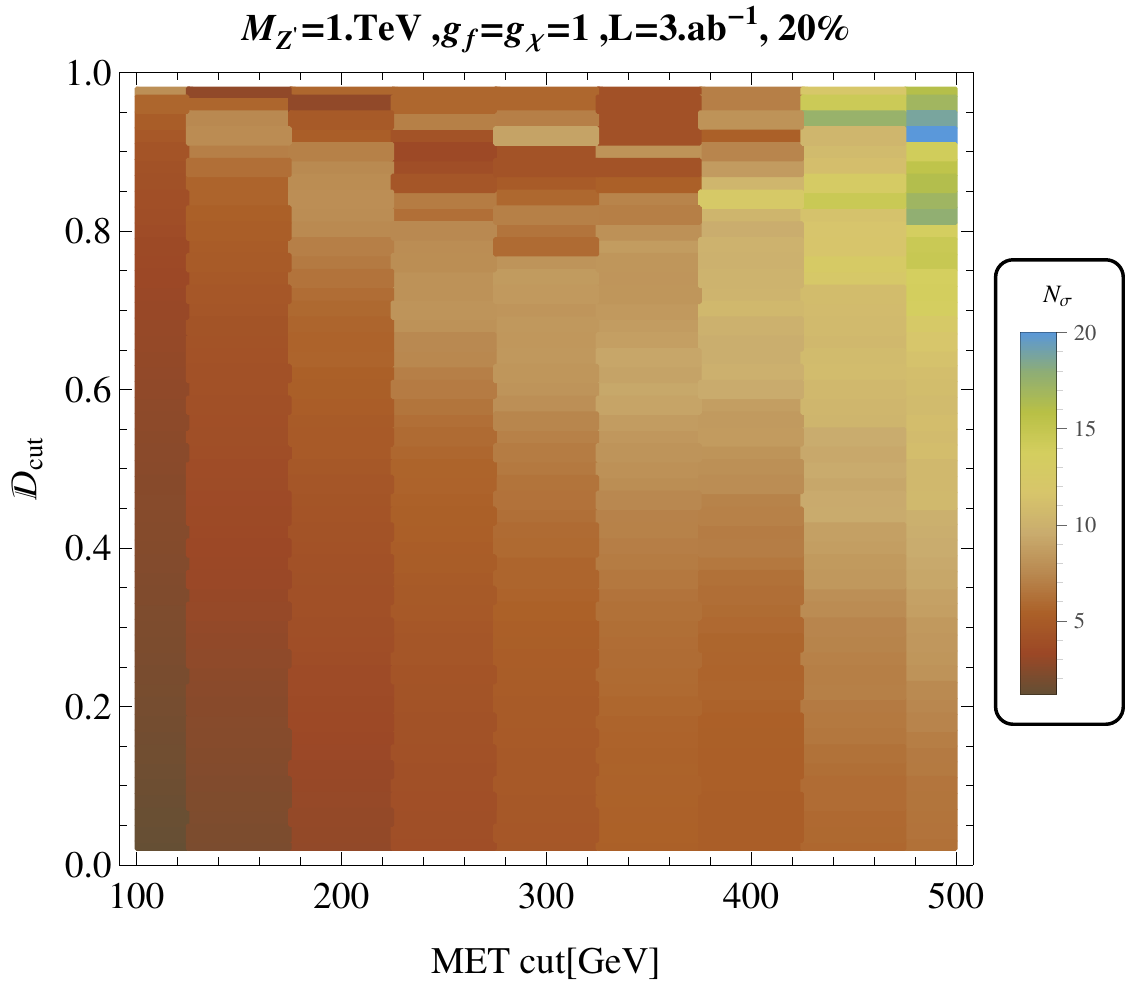}}
  \caption{The reach($5\sigma$) for the cut-off scale $\Lambda$ is shown at the upper and middle rows for various choices of cuts on $\mathcal{D}_{LF}$ (vertical axis) and $\met$ (horizontal axis), for 3 ab$^{-1}$ of integrated luminosity. At the upper row we show the results for the $\tilde{D}1$ operator, at the middle row the D5 operator. At the lower row, the statistical significance for the discovery of an 1 TeV $\zp$ with couplings $g_f=g_\chi=1$ is shown. At the left(right) panels the systematic uncertainty is 5(20)\%.} 

  \label{Lambdareach}
\end{figure}


\begin{figure}[!ht]
\centering
\includegraphics[width=0.6\textwidth]{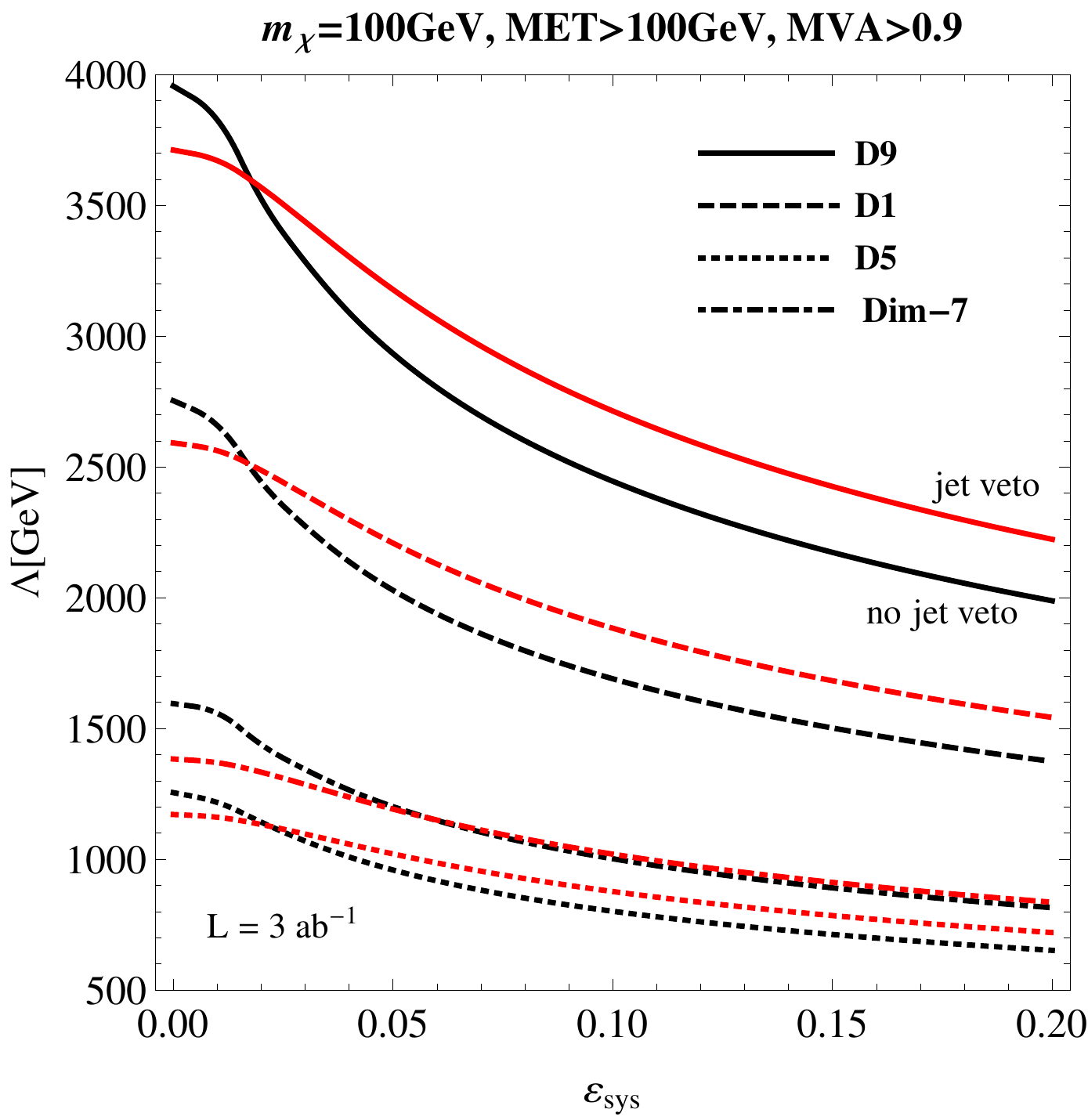}
\caption{\textbf{Reach in } $\mathbf{\Lambda}$  \textbf{for EFT operators}: The reach in $\Lambda$ for the EFT operators at 3000 fb$^{-1}$ of integrated luminosity. The red curves display the case when jet veto is applied, while the black curves display the case when no jet veto is applied. The mass of the DM candidate is fixed at 100 GeV. The cuts applied are $\met \, > \, 100$ GeV and $\mathcal{D}_{LF} \, > \, 0.9$.}
\label{EFT_final_results}
\end{figure}

In the next Section, we go on to give our main results for the EFT operators.

\section{Results for EFT Operators} \label{EFTresults}

As described previously, the reach for $\Lambda$ depends on a judicious choice of $\met$ cut and a cut on the likelihood function $\mathcal{D}_{LF}$. We display the discovery reach ($5\sigma$) for a variety of luminosities and choices of systematics. The results for the operator $\tilde{\hbox{D}}1$, with a jeto veto, are shown in the upper row of Fig.~\ref{Lambdareach}. The reach for the cut-off scale $\Lambda$ is shown for various choices of cuts on $\mathcal{D}_{LF}$ (vertical axis) and $\met$ (horizontal axis), for different luminosities and systematics. The reach for luminosity 300 fb$^{-1}$, for systematics of $5\%$ ($20\%$), is  1857 (1500) GeV. The reach for luminosity 3000 fb$^{-1}$, for systematics of $5\%$ ($20\%$), is 2209 (1542) GeV. As is clear from the figure, the maximum reach is obtained for $\met \, > \, 100$ GeV, $\mathcal{D}_{LF} \, > \, 0.9$. We observed very similar behavior concerning operators D9, dim-5 and dim-7. The reach for the dim-5 operator is limited to a few tens of GeV even for low systematics and high luminosities. We summarize in Table~(\ref{resultsEFT}) the results for all operators but dim-5.

The results for the operator D5 are shown in the middle row of Fig.~\ref{Lambdareach}. The results for operator D8 are similar. The reach for luminosity 300 fb$^{-1}$, for systematics of $5\%$ ($20\%$), is 843 (695) GeV. The reach for luminosity 3000 fb$^{-1}$, for systematics of $5\%$ ($20\%$), is 1021 (720) GeV. In this case, however, we see that a harder $\met$ cut degrades the achievable reach of the LHC compared to the other operators. This is, again, due the larger similarity between the shapes of the kinematic distributions of D5 (and D8) operator and the backgrounds as can be seen in Figs.~(\ref{fig:sub2}), (\ref{fig:sub1}) and (\ref{fig:sub3}).



The final results at 3000 fb$^{-1}$ for the reach in $\Lambda$ for all the EFT operators are displayed in Fig.~\ref{EFT_final_results}. The reach in $\Lambda$ for the EFT operators is shown at 3000 fb$^{-1}$ of integrated luminosity. The red curves display the case when a jet veto is applied, while the black curves display the case when no jet veto is required. The mass of the DM candidate is fixed at 100 GeV throughout all our analyses. The cuts applied are $\met \, > \, 100$ GeV and $\mathcal{D}_{LF} \, > \, 0.9$. The solid, long-dashed, short-dashed, and dot-dashed curves denote the limits for D9, $\tilde{\hbox{D}}1$, D5, and the dim-7 operators, respectively. The D8 curves are very close to the D5 ones.

First of all, we see that including systematic uncertainties is a really necessary ingredient to get realistic estimates for the reach of the LHC for dark matter. Keeping systematics under control might improve the reach discovery by a factor of two roughly. For example, in the D9 case, a 20\% systematics in the backgrounds reduces the reach to $\sim 2200$ GeV, while keeping a 5\% level or less increases the reach to $\sim 3500$ GeV or farthest. 

Adopting a jet veto also helps to get a larger reach for dark matter searches in the mono-$Z$ channel if the systematics are larger than approximately 3-5\%. 

%
\begin{table}
\caption{Results for EFT operators and the leptophobic dark $\zp$ portal for three different selection cuts, integrated luminosities of 0.3 and 3 ab$^{-1}$, and systematic uncertainties of 5\% and 20\% in the total background rate. The last row shows the effect of including a jet veto in the analysis. The blank entries in the last column indicate that a 5$\sigma$ discovery cannot be reached for that luminosity and level of systematics.}
\label{resultsEFT}
\begin{tabular}{ l|c|c|rrrrrr } 
\hline 
CUTS & L(ab$^{-1}$) & sys(\%) & ~~~~~~$\tilde{\hbox{D}}1$ & ~~~~~~D5 & ~~~~~~D8 & ~~~~~~D9 & ~~~~~~Dim-7 & ~~~~~~$M_{\zp}$\\
\hline
\multirow{2}{*}{$\met > 100$ GeV} & \multirow{2}{*}{0.3} & 5 & 702 & 404 & 398 & 994 & 400 & 327\\
\cline{3-9}
                                                         & & 20 & 470 & 270 & 266 & 665 & 268 & --\\
\cline{2-9}
                                  & \multirow{2}{*}{3}   & 5 & 702 & 404 & 398 & 996 & 401 & 328\\
\cline{3-9}
                                                         & & 20 & 470 & 270 & 266 & 666 & 268 & --\\
\hline 
\multirow{2}{*}{$\met > 100$ GeV + $\mathcal{D}_{LF}>0.9$} & 
\multirow{2}{*}{0.3} & 5 & 1864 & 868 & 859 & 2686 & 1096 & 675\\ 
\cline{3-9}
                     & & 20 & 1363 & 645 & 635 & 1972 & 808 & 675\\
\cline{2-9} &
\multirow{2}{*}{3} & 5 & 2030 & 959 & 859 & 2944 & 1201 & 1690\\ 
\cline{3-9}
                   & & 20 & 1374 & 652 & 635 & 1988 & 815 & 1010\\
\hline 
\multirow{2}{*}{$\met > 100$ GeV + $\mathcal{D}_{LF}>0.9$ + jet veto} & 
\multirow{2}{*}{0.3} & 5 & 1857 & 843 & 835 & 2663 & 994 & 1233\\
\cline{3-9}
                     & & 20 & 1500 & 695 & 684 & 2159 & 810 & 906\\
\cline{2-9} &
\multirow{2}{*}{3} & 5 & 2209 & 1021 & 1005 & 3178 & 1192 & 1485\\
\cline{3-9} 
                     & & 20 & 1542 & 720 & 707 & 2224 & 836 & 937\\
\hline \end{tabular}
\end{table}

\subsection{Comparative Performances of MVA and MET Cut for EFT Operators}
\begin{figure}[!ht]
  \centering
  {\includegraphics[width=.32\textwidth]{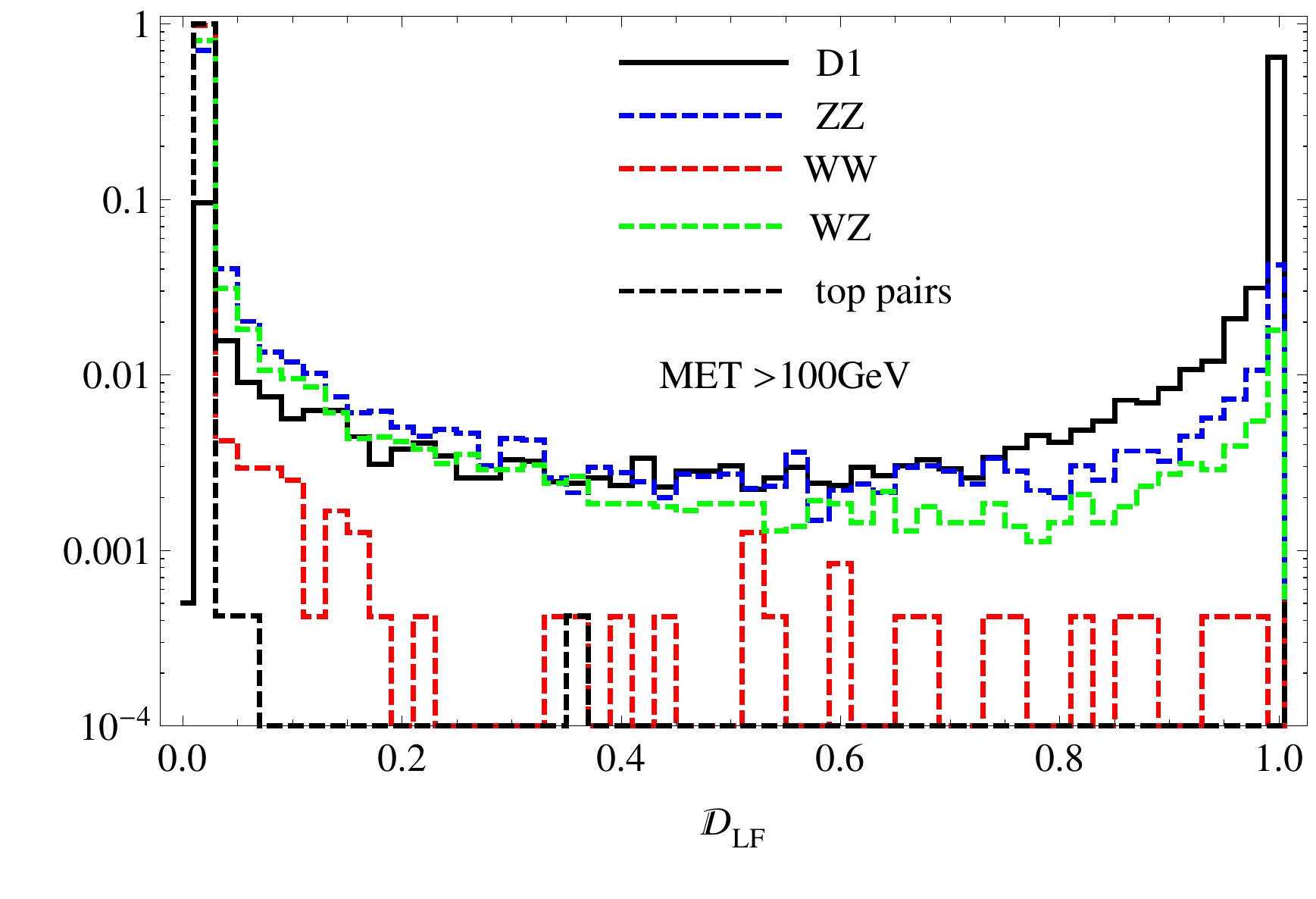}}\quad
  {\includegraphics[width=.32\textwidth]{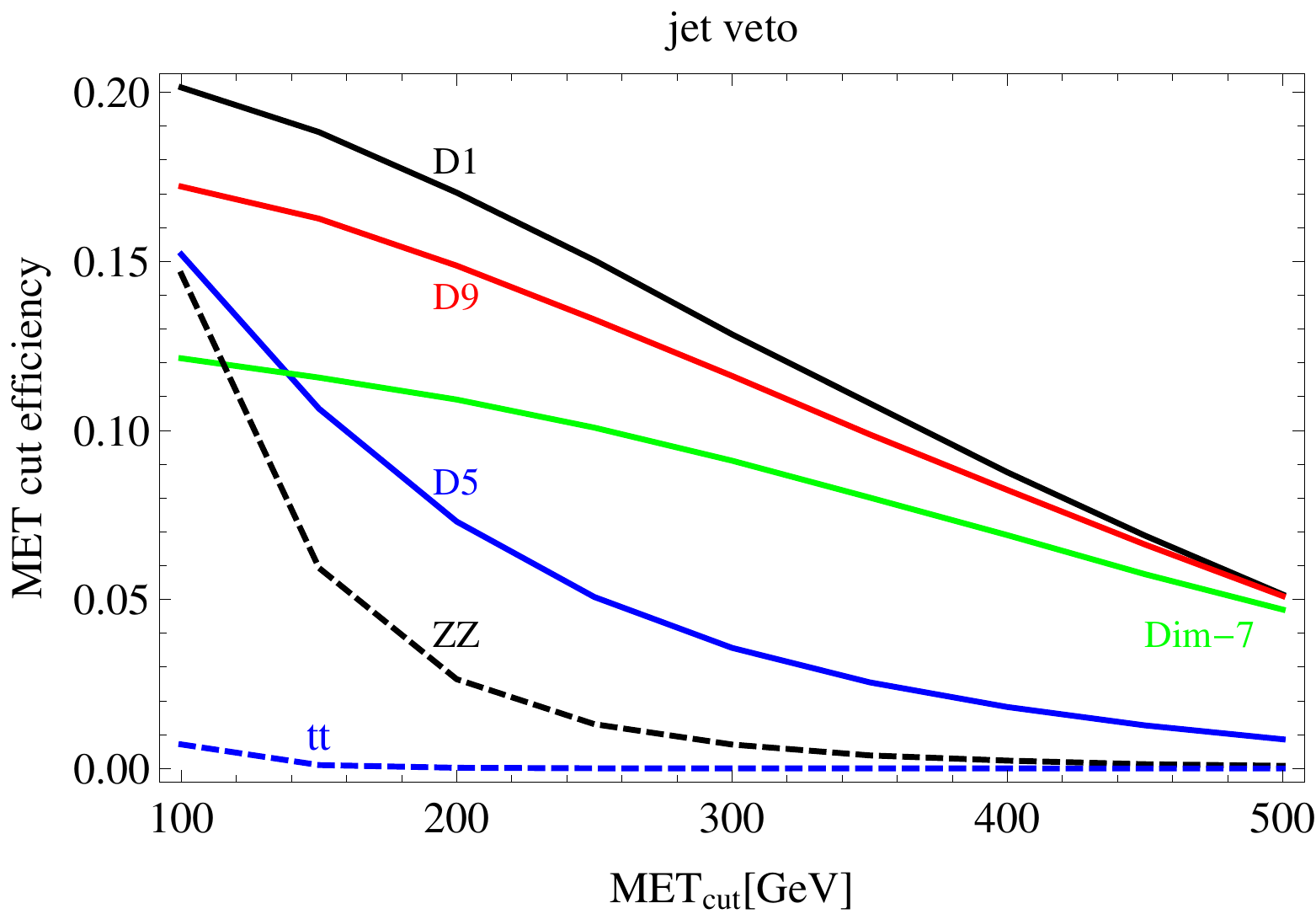}}\quad
  {\includegraphics[width=.32\textwidth]{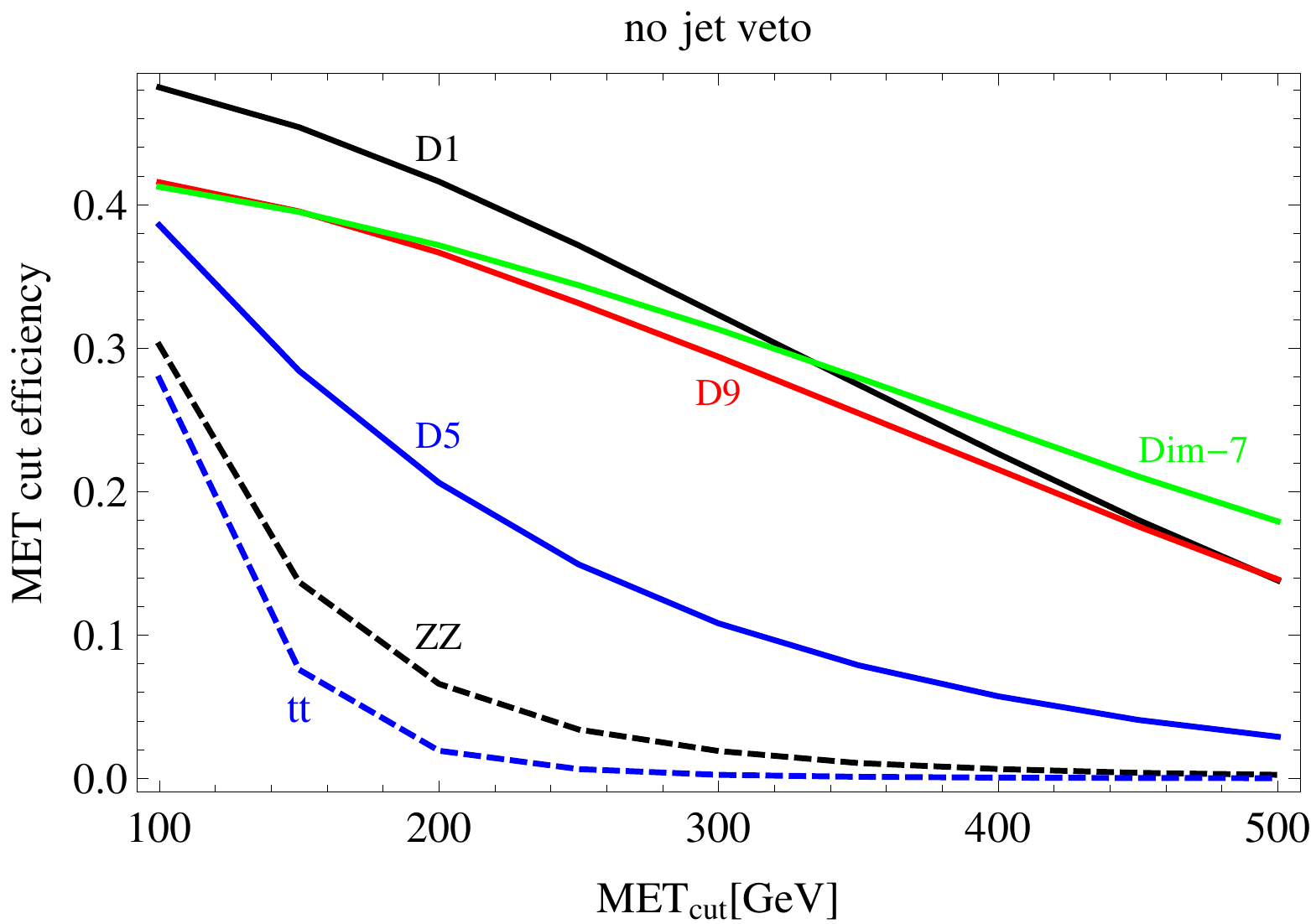}}\quad
  \caption{In the left panel, the MVA discriminant distribution for the $\tilde{\hbox{D}}1$ operator (solid line) and all the relevant backgrounds (dashed lines) for a mono-$Z$ search. In the central(right) panel we display the cut efficiency with just a MET cut of 100 GeV with(without) a jet veto for all the EFT operators, but dim-5, and for $ZZ$ and top pair backgrounds.}
  \label{cuteff}
\end{figure}

 The lesser sensitivity to systematics is the result of higher $S/B$ ratios when we better classify the events according to the MVA discriminant. We checked that the signal to background ratio increases of an order of magnitude after imposing $\mathcal{D}_{LF} >0.8$ compared to the MET cut analysis alone, for all MET cuts, reaching $S/B\sim 7$ for a 100 GeV MET cut and $\mathcal{D}_{LF} >0.9$ with and without a jet veto. 

In the 13 TeV LHC, top quark pair production is not as harmless as in 7 or 8 TeV runs concerning the mono-$Z$ channel. In this respect the combination of a MVA discriminant and a jet veto can be used to eliminate background events not involving a SM $Z$ boson. This can be seen in Fig.~(\ref{cuteff}). First, in the left panel, note that the MVA distributions of $t\overline{t}$ and $WW$ backgrounds are highly concentrated near $\mathcal{D}_{LF}=0$, while those $ZZ$ and $ZW$ are more widely distributed. Demanding $\mathcal{D}_{LF} > 0.5$ gets rid of all top events and the majority of $WW$ backgrounds. In the central(right) panel we show the cut efficiency for a MET cut only with(without) the jet veto for $\tilde{\hbox{D}}1$, D5, D9, and dim-7 operators, and the $ZZ$ and top pair backgrounds. Again, the top backgrounds can be more efficiently eliminated by tagging the hard bottom jets from top decays.

\begin{figure}[!ht]
  \centering
  {\includegraphics[width=.45\textwidth]{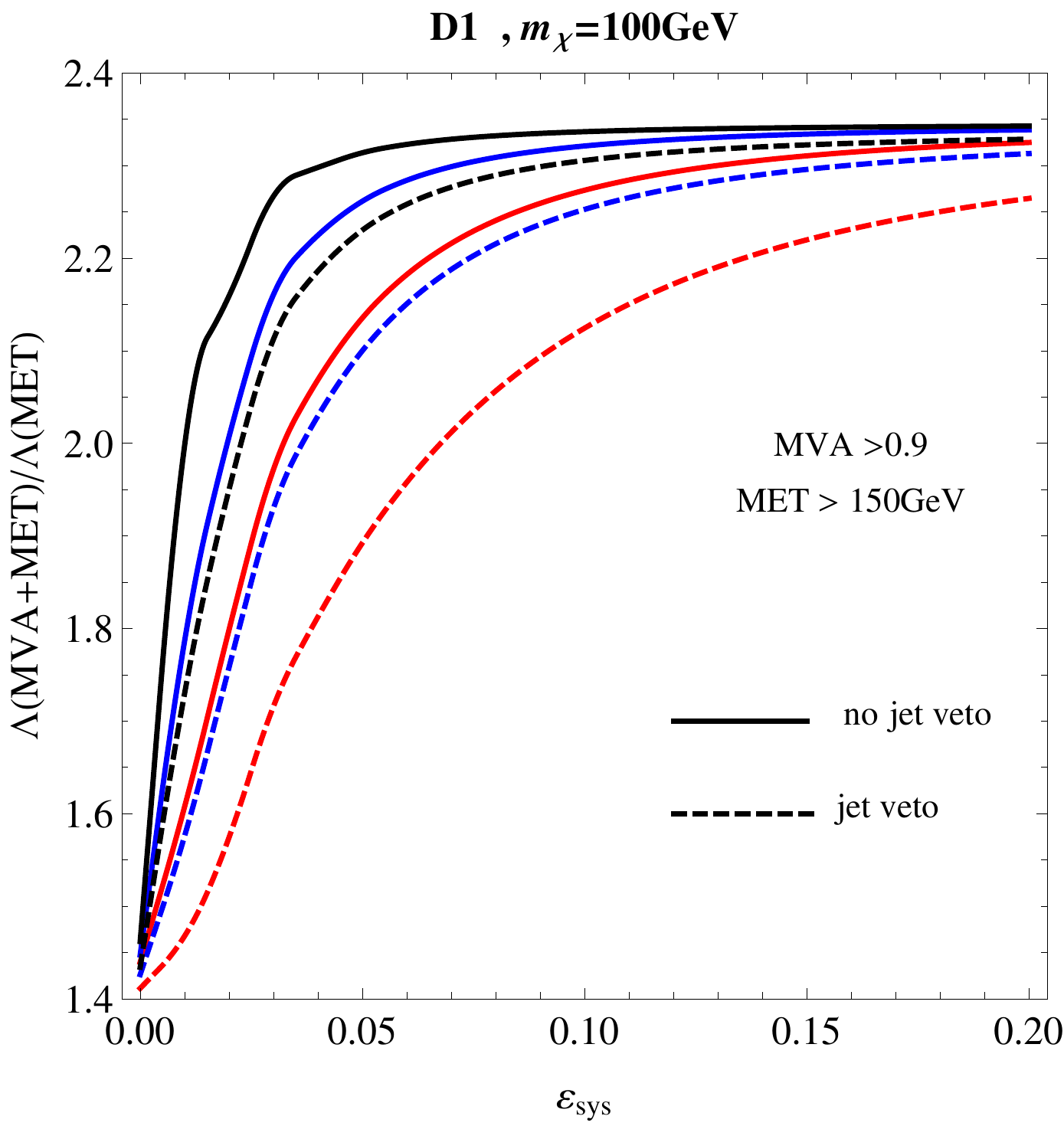}}\quad
  {\includegraphics[width=.45\textwidth]{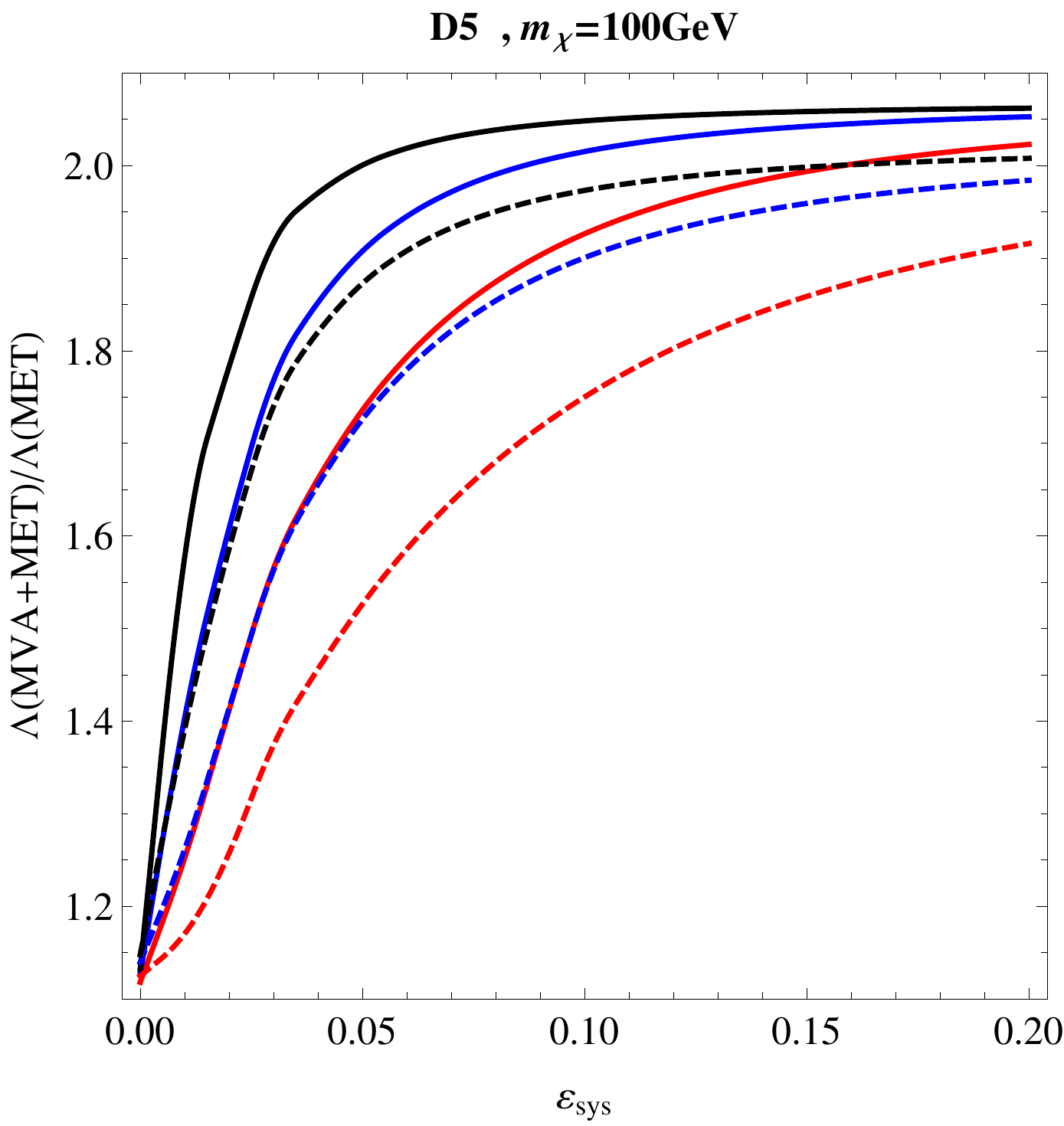}}\\
  {\includegraphics[width=.45\textwidth]{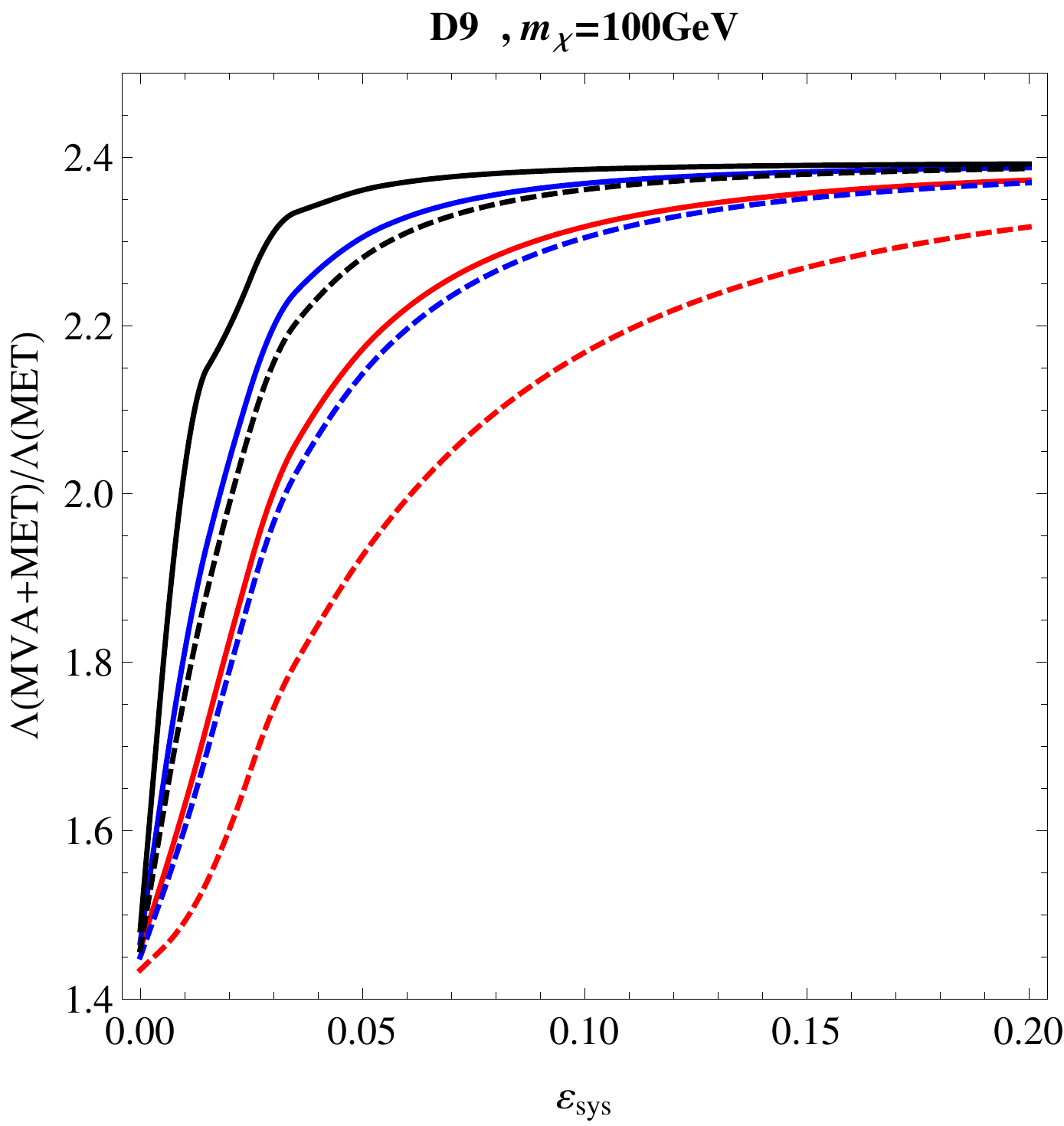}}\quad
  {\includegraphics[width=.45\textwidth]{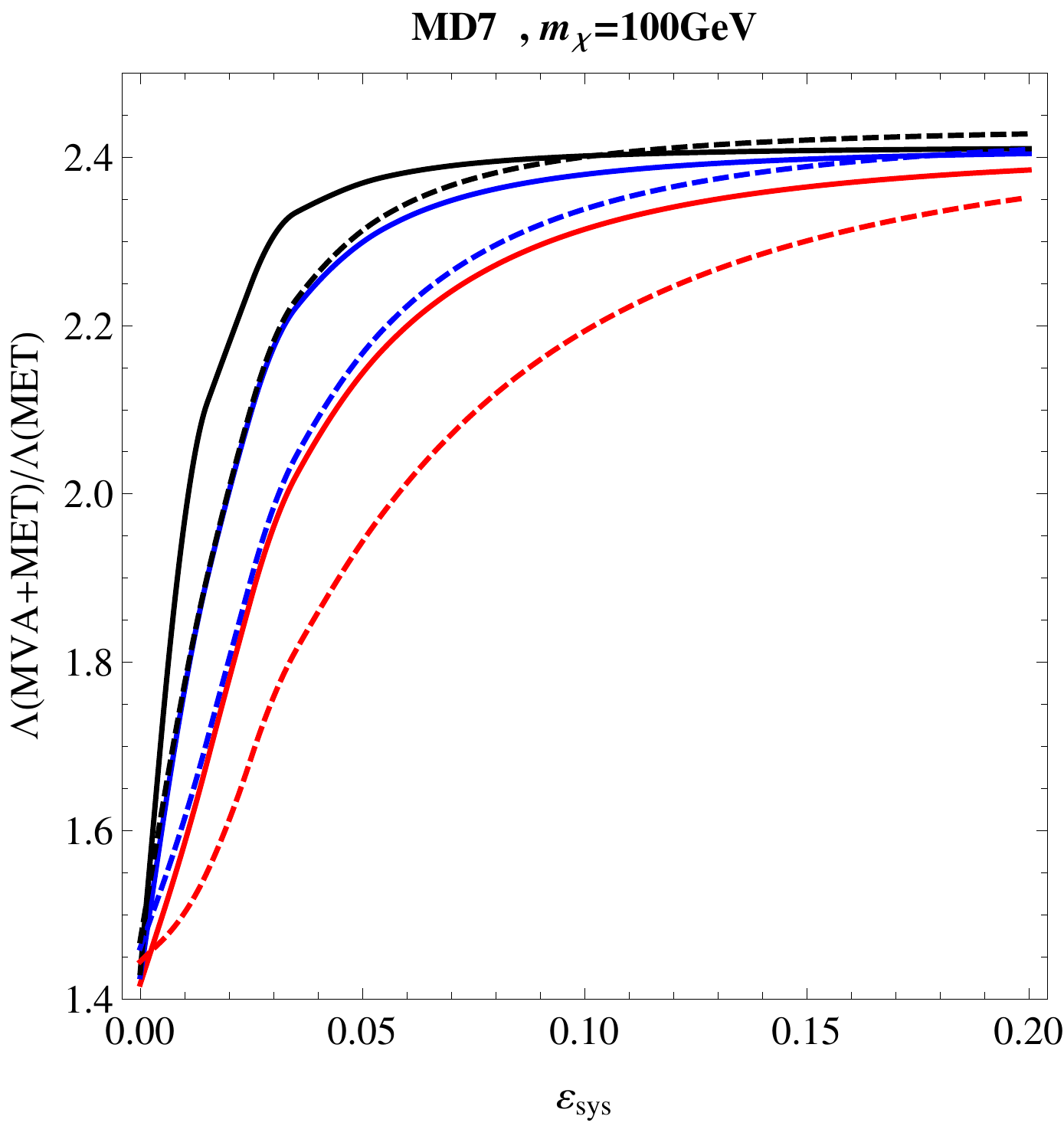}}
  \caption{The ratio of the reach in $\Lambda$, as a function of systematics, when the $\met$ cut is combined with MVA, compared to when only the $\met$ cut is applied, for $\tilde{\hbox{D}}1$, D5, D9, and dim-7(MD7) operators. The dashed(solid) lines show the ratios when we do(not) apply a jet veto. The lower red curves were obtained fixing the luminsoity at 300 fb$^{-1}$, the central blue ones at 1000 fb$^{-1}$, and the upper black curves at 3000 fb$^{-1}$.}
  \label{MVAmetComparison}
\end{figure}

Let us discuss the advantage of using the multivariate analysis that we have described, when compared to a simple $\met$ cut. In Fig.~\ref{MVAmetComparison}, we show the ratio of the reach in $\Lambda$ when the $\met$ cut is combined with MVA, compared to when only the $\met$ cut is applied. With no systematics, the reach in $\Lambda$ with MVA is approximately $1.2 \, \, - \,\, 1.4$ times larger than the case with only $\met$ cut. This ratio increases dramatically with increasing systematics, reaching $2.0 \,\,-\,\,2.4$ for $\sim \, 5\%$ systematics, after which it becomes stable for a wide range of integrated luminosities from 300 to 3000 fb$^{-1}$. This clearly shows the advantage of using the MVA for higher systematics.

\section{Results for the leptophobic dark $\zp$ portal model} \label{ZPresults}

Applying the multivariate analysis to the same kinematic observables in this case, we obtain the likelihood function for different masses of $Z^{\prime}$. We will take $g_f=g_\chi=1$ and $m_\chi=100$ GeV in what follows. 

 In the last row of Fig.(~\ref{Lambdareach}), the significance is shown for the mediator mass $m_{Z^{\prime}} \, = \, 1$ TeV, for an integrated luminosity of 3 ab$^{-1}$ and 5 and 20\% systematics. 
 
\begin{figure}[!htp]
\centering
\includegraphics[width=3.3in]{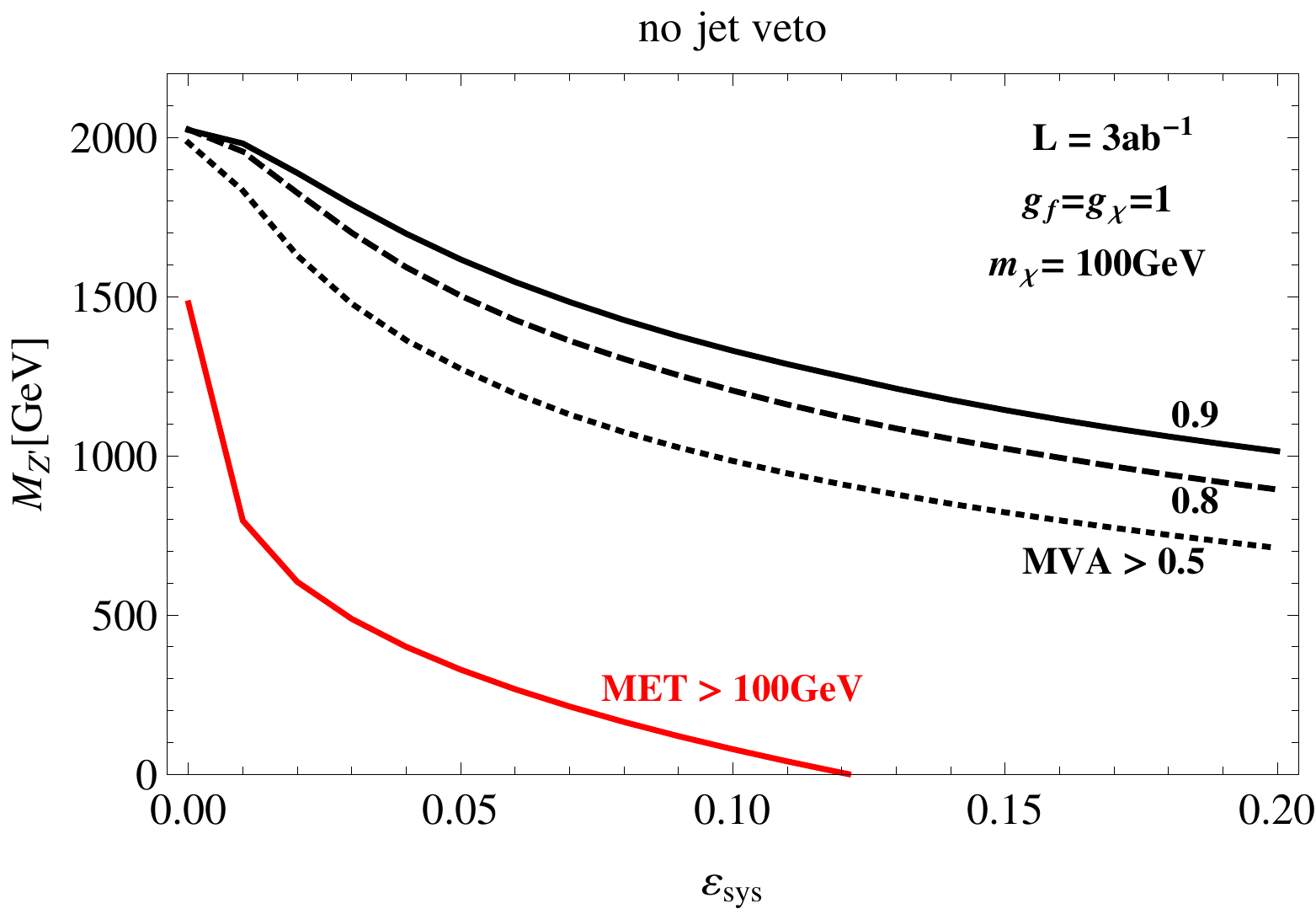}
\includegraphics[width=3.3in]{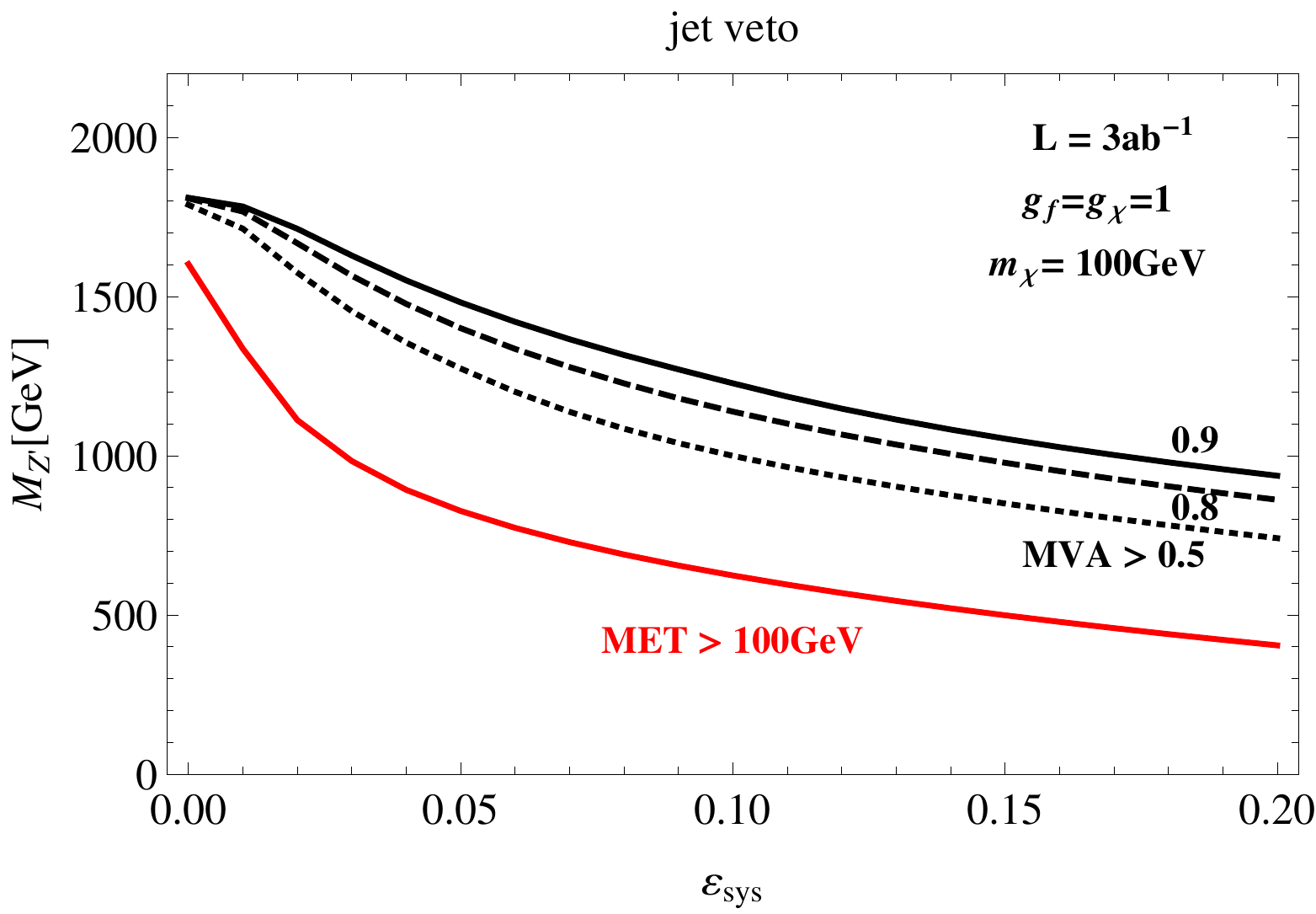}
\caption{\textbf{Reach in } $\mathbf{M_{Z^{\prime}}}$: The $5\sigma$ reach in $M_Z^\prime$ for $\met \, > \, 100$ GeV cut only (red curve) and $\met$ cut combined with different cuts on the likelihood function $\mathcal{D}_{LF}$: 0.9 (black solid), 0.8 (black dashed), and 0.5 (black dotted), at 3000 fb$^{-1}$  of integrated luminosity. The dark matter mass is fixed at 100 GeV, and all couplings are set to one. At the left(right) panel we show the results without(with) a jet veto.}
\label{Zprime_final_results}
\end{figure}

In Fig.~\ref{Zprime_final_results}, the $5\sigma$ reach in $m_Z^\prime$ for $\met \, > \, 100$ GeV cut only and $\met$ cut combined with different cuts on the likelihood function $\mathcal{D}_{LF}$ are shown, at 3 ab$^{-1}$  of integrated luminosity. It is clear from the results that for no systematics, the reach with MVA is around 2 TeV, while the reach with $\met$ cut is around 1.5 TeV.  For increasing systematics, there is rapid degeneration of this reach for the case when only $\met$ cut is used. For the MVA analysis, the degeneration is much more gradual, and even for high systematics, the reach is still above 1 TeV. 

Contrary to the case of EFT operators, imposing no jet veto on the events is advantageous for all systematics if we combine MVA and MET cuts. However, for a MET cut alone, the reach is much more depleted in comparison to the jet veto case. A $5\sigma$ signal cannot be reached even for 3 ab$^{-1}$ if $\varepsilon_{sys}>0.12$ as we see in the left panel of Fig.~(\ref{Zprime_final_results}). We conclude that a jet veto removes more signal and background events, but with a higher signal to background ratio.

 As in the case of EFT models, the sensitivity to systematics is greatly reduced as a result of higher $S/B$ ratios when we better classify the events according to the MVA discriminant. We also checked, for these $\zp$ models, that the signal to background ratio increases of an order of magnitude after imposing $\mathcal{D}_{LF} >0.8$ compared to the MET cut analysis alone without a jet veto.


%

\section{Conclusions} \label{conclusions}

In this paper, we have investigated the discovery prospects of DM effective operators and simplified models with vector mediators at Run II of the LHC. We have paid particular attention to the question of systematic uncertainties. To that end, we have based our study on a multivariate analysis in the mono-$Z$ channel. 

Our main results are the following. We have seen that the reach obtained using a multivariate discriminant is typically at least twice as high as that obtained from a simple cut and count analysis, once systematic uncertainties in the total background normalization higher than a few percent are taken into account. This is clear from Fig.~\ref{MVAmetComparison}. With no systematics, the reach in $\Lambda$ with the multivariate analysis is approximately $1.2 \, \, - \,\, 1.4$ times larger than the case with only $\met$ cut. This ratio increases dramatically with increasing systematics, reaching $2.0 \,\,-\,\,2.4$ for $\sim \, 5\%$ systematics, after which it becomes stable. This clearly shows the advantage of using the MVA for higher systematics.

Moreover, the reach is much more stable against degradation due to systematics, both for the EFT analysis as well as for the $\zp$ analysis. This is clear from Fig.~\ref{Zprime_final_results}, where we show the results of the $\zp$ analysis. For no systematics, the reach in $m_{\zp}$ with multivariate analysis is around 2 TeV, while the reach with only $\met$ cut is around 1.5 TeV.  For increasing systematics, there is rapid degeneration of the reach for the case when only $\met$ cut is used. For the multivariate analysis, the degeneration is much more gradual, and even for high systematics, the reach is still above 1 TeV. 

We also investigated the impact of including a jet veto to suppress backgrounds further. We found that a jet veto improves the reaches for the EFT as long as the level of systematics exceedes 3--5\%. For the leptophobic $\zp$ simplified model, requiring no jet veto provides better results when we perform the multivariate analysis, but it is important if just a MET cut is used.

Our reaches are the following. For DM-quark effective operators with scalar, vector, and tensor couplings and DM mass 100 GeV, the 5$\sigma$ reaches in the DM interaction scale $\Lambda$ keeping the level of systematics at 5\% with 300 (3000) fb$^{-1}$ of integrated luminosity are 1864 (2209) GeV, 959 (1021) GeV, and 2686 (3178) GeV, respectively. For simplified models with leptophobic vector mediators, the 5$\sigma$ reach for the mass of the mediator is 1233 (1485) GeV with 5\% systematics and 300 (3000) fb$^{-1}$ of data. 

\section{Acknowledgments}

A.A. is supported by Fundac\~ao de Amparo \`a Pesquisa do Estado de S\~ao Paulo (FAPESP) grant 2013/22079-8 and Conselho Nacional de Desenvolvimento Cient\'ifico e Tecnol\'ogico (CNPq), grant 307098/2014-1. K.S. is supported by NASA Astrophysics Theory Grant NNH12ZDA001N. K.S. would like to thank the CETUP* 2015 Dark Matter Conference in South Dakota for providing a stimulating atmosphere where part of this work was completed. A.A. would like to thank Farinaldo Queiroz for helpful discussions in the earlier stages of this work. The authors would especially like to thank Tim Tait for very helpful discussions and for carefully reading a previous version of the draft.

\appendix

\section{Details of the significance computation}

A comprehensive study of various methods for the calculation of statistical significances has been presented in~\cite{Cousins}. In this work, the problem of incorporating systematic uncertainties in the background normalization for a Poisson process is addressed
and it is found that three most widely used significance metrics perform similarly in many situations concerning the relative number of signal and background events and the level of systematics.

\begin{figure}[!hbp]
\centering
\includegraphics[width=3.in]{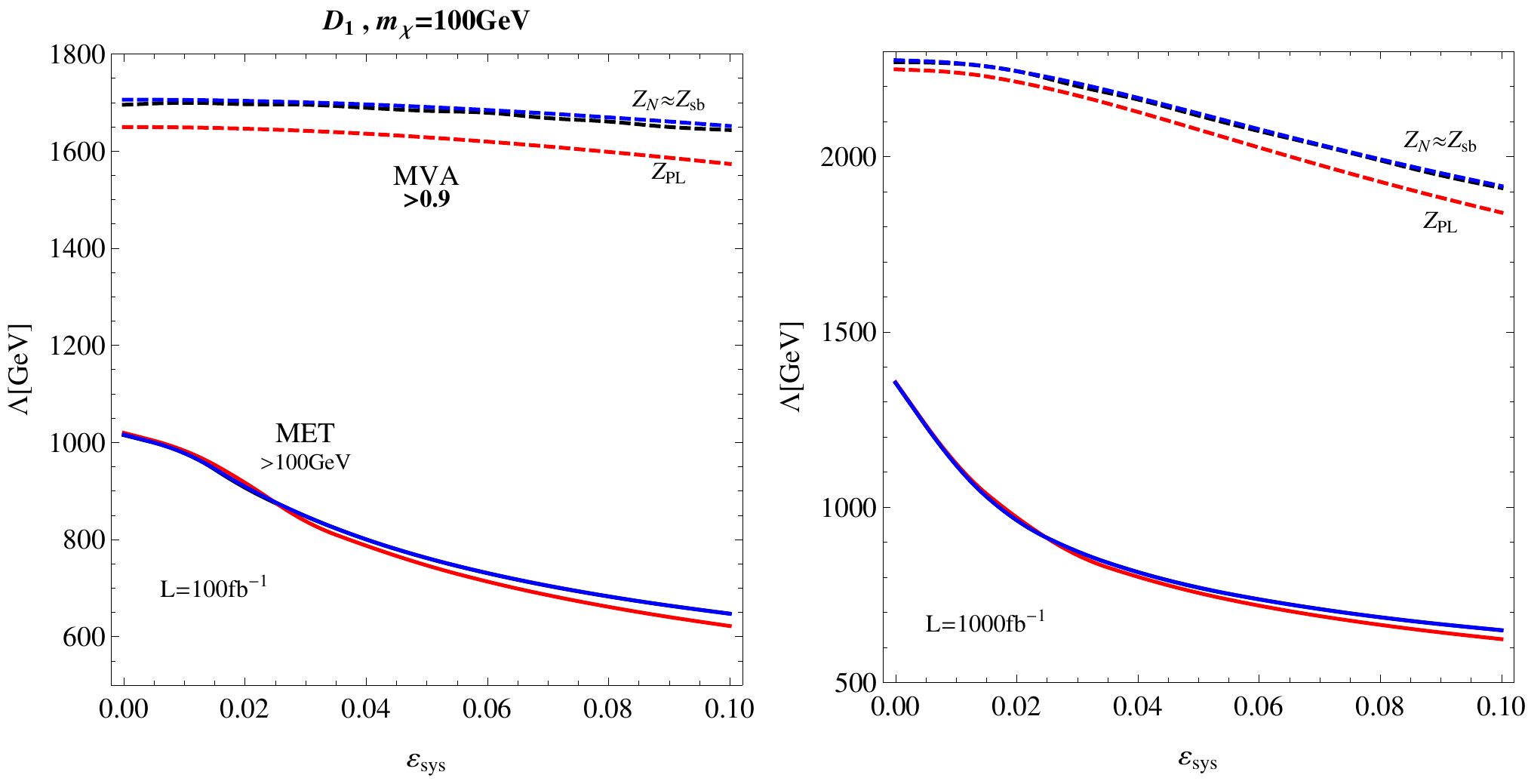}
\includegraphics[width=3.in]{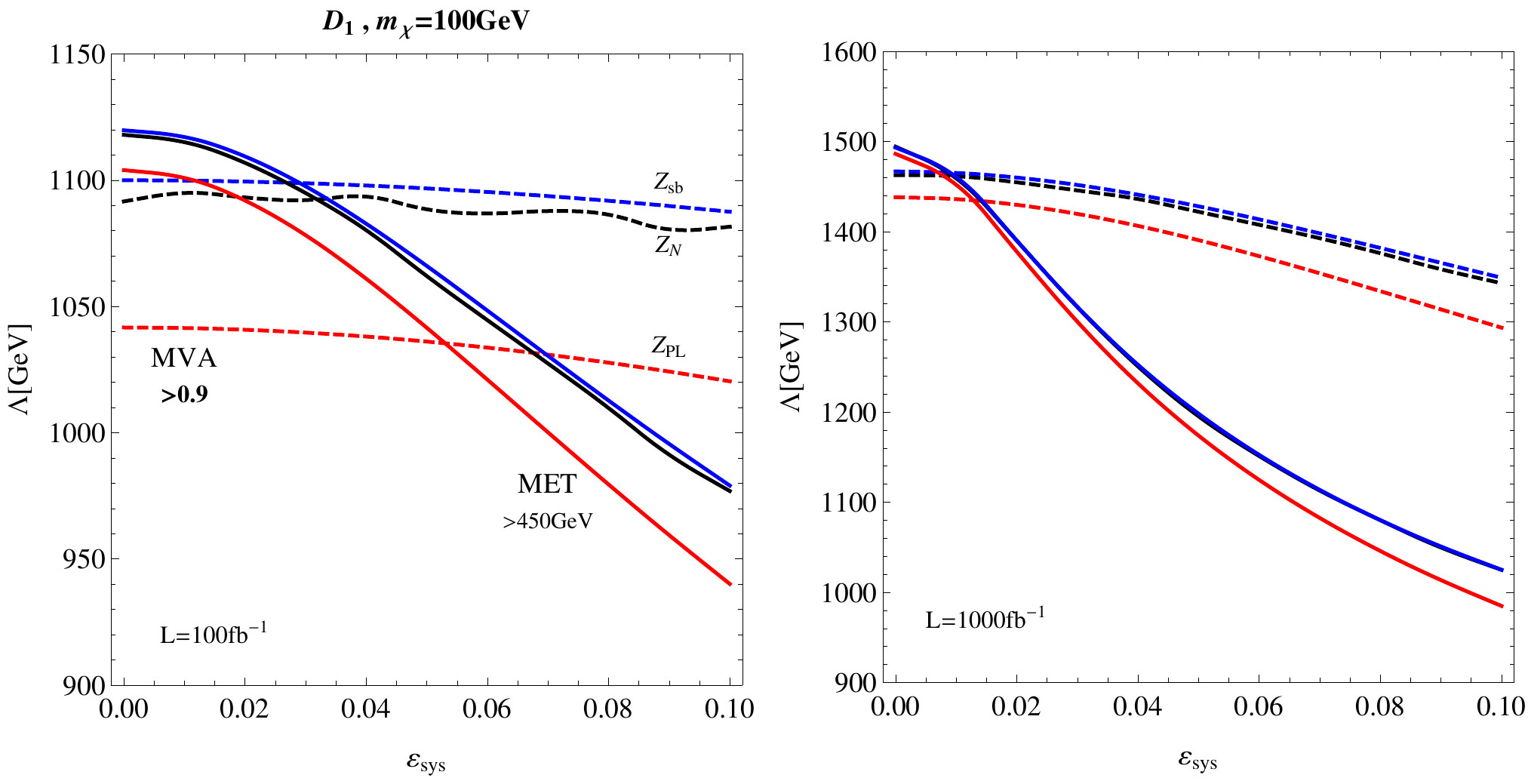}
\caption{Comparison of significance calculation methods. The solid(dashed) lines represent the reach in $\Lambda$ for the scalar $\tilde{\hbox{D}}$1 operator for $m_\chi=100$ GeV with MET(MET+MVA) cuts as a function of the systematic uncertainties $\varepsilon_{sys}$ for 100 and 1000 fb$^{-1}$. The upper, middle and lower dashed(solid) lines correspond to $Z_{sb}$, $Z_N$, and $Z_{PL}$ metrics respectively. For solid lines we apply only a 100 GeV or 450 GeV MET cut, while for the dashed ones we also impose the MVA cut $\mathcal{D}_{LF} > 0.9$.}
\label{Zcompare}
\end{figure}

The three metrics are computed as follows:

\begin{itemize}
\item[\underline{$Z_{sb}$}] This is the naive and most simple way to incorporate systematic uncertainties in the calculation of the significances for $S$ signal events and $B$ background events in a Poisson process for a given integrated luminosity
$$
Z_{sb}=\frac{S}{\sqrt{B+\sigma_{sys}^2}}
$$

In all cases, we assume that the systematic uncertainty in the total background normalization is proportional to the number of background events, $\sigma_{sys}=\varepsilon_{sys} B$. This is simple and fast, but as in~\cite{Cousins}, we found it somewhat overestimates the reach as we show in Fig.~(\ref{Zcompare}), the upper blue lines.

\item[\underline{$Z_N$}] The the Bayesian-frequentist hybrid recipe to the estimation of the systematics impact on the significance. Assuming that systematic errors are normally distributed we marginalize over the systematic errors to obtain the p-value 
$$
p=\sum_{k=S+B}^{+\infty} \int_{-\infty}^{+\infty} \frac{e^{-B(1+z\varepsilon_{sys})}}{k!}\left[B(1+z\varepsilon_{sys})\right]^k\times \frac{e^{-\frac{z^2}{2}}}{\sqrt{2\pi}}\, dz
$$
and the significance is computed as $Z_N=\Phi^{-1}(1-p)$, where $\Phi(z)$ is the cumulative distribution function of the standard normal distribution.

This method demands a much larger computation effort. As we show in Fig.~(\ref{Zcompare}), the middle red lines, its performance is almost identical to the naive $Z_{sb}$ metric though, despite the reach in $\Lambda$ is always shorter than $Z_{sb}$. In the right panel we see that in the regime of a low number of events, $Z_N$ might present numerical instabilities.

\item[\underline{$Z_{PL}$}] The Profile Likelihood method originally proposed in~\cite{LiMa} in astrophysical searches with subsidiary measurements of the background adapted to a high energy experiment where the systematics is a fraction of background events
$$
Z_{PL}=\sqrt{2}\left\{(S+B)\ln\left[\left(1+\frac{B}{\sigma_{sys}^2}\right)\frac{S+B}{S+B+B^2/\sigma_{sys}^2}\right]
+\frac{B^2}{\sigma_{sys}^2}\left[\left(1+\frac{\sigma_{sys}^2}{B}\right)\frac{B^2/\sigma_{sys}^2}{S+B+B^2/\sigma_{sys}^2}\right]\right\}^{\frac{1}{2}}
$$

Among the three metrics this is the most conservative and reliable, and it is as simple and fast to compute as $Z_{sb}$~\cite{Cousins}. Moreover, its performance is very close to the consistent frequentist approach for tests of the ratio of Poisson means implemented in \texttt{ROOT}~\cite{root}, for example. The $Z_{PL}$ results correspond to the lower black lines in Fig.~(\ref{Zcompare}).

\end{itemize}

In order to choose judiciously among these three significances metrics, a performance comparison for our benchmark point in the case of determining the LHC reach for the scalar operator $\tilde{\hbox{D}}$1 was carried out. In the Fig.~(\ref{Zcompare}) we show the reach in $\Lambda$ using $Z_{sb}$ (upper blue), $Z_N$ (middle red), and $Z_{PL}$ (lower black) in the multivariate analysis (dashed lines) and the simple cut-and-count analysis (solid lines). In the left(right) panels we assume a 100(1000) fb$^{-1}$ of integrated luminosity for $\mathcal{D}_{LF}>0.9$ and a MET cut of 100 GeV and 450 GeV.

The Profile Likelihood metric $Z_{PL}$ is the most conservative one, thus confirming the findings of~\cite{Cousins}. The reaches, however, are similar, especially comparing $Z_{sb}$ and $Z_N$ in the regime of a higher number of signal and background events. As the number of events decrease, as in the case of a hard MET cut of 450 GeV, the overestimation of $Z_{sb}$ and $Z_N$ gets more noticeable. Also, numerical instabilities appear in the computation of $Z_N$ and a greater care is necessary when using this method.

Using the multivariate discriminant to better separate signal and backgrounds increases the S/B ratio compared to a simple MET cut.
In this regime (dashed lines), $Z_{sb}$ and $Z_N$ also overestimate the reach compared to $Z_{PL}$, but the performances are very close with a MET cut only (solid lines), that is it, for smaller S/B ratios.

\end{document}